\documentclass[twocolumn,aps,prb,showpacs,amsmath,superscriptaddress]{revtex4}
        \usepackage{amsmath}
        \usepackage{amssymb}
        \usepackage{subfigure}
        \usepackage{natbib}
        \usepackage{epsfig}
        \usepackage{amsfonts}
        \usepackage{mathrsfs}
        \usepackage{ulem}
        \usepackage{color}
        \usepackage{bm}
        \usepackage[toc,page,title,titletoc,header]{appendix}
        \usepackage{CJK}
        \usepackage{epstopdf}
    \usepackage{mathrsfs}
    \usepackage{float}
    \usepackage{tikz-cd}
    \usepackage{caption}
    \usepackage{graphicx, subfig}
    \usepackage{dcolumn}
    \usepackage{bm}
\usepackage{multirow}
\usepackage{comment}
\usepackage{indentfirst}
   \begin{document}
      \widetext
      \title{Topological classification of defects in non-Hermitian systems}
      \author{Chun-Hui Liu}
      \email{liuchunhui@iphy.ac.cn}
         \affiliation{Beijing National Laboratory for Condensed Matter Physics, Institute of Physics, Chinese Academy of Sciences, Beijing 100190, China}
\affiliation{School of Physical Sciences, University of Chinese Academy of Sciences, Beijing 100049, China}
\author{Shu Chen}
\email{schen@iphy.ac.cn}
\affiliation{Beijing National Laboratory for Condensed Matter Physics, Institute of Physics, Chinese Academy of Sciences, Beijing 100190, China}
\affiliation{School of Physical Sciences, University of Chinese Academy of Sciences, Beijing 100049, China}
\affiliation{Yangtze River Delta Physics Research Center, Liyang, Jiangsu 213300, China}

          \begin{abstract}
           \par
             We classify topological defects in non-Hermitian systems with point gap, real gap and imaginary gap for all the Bernard-LeClair classes or generalized Bernard-LeClair classes in all dimensions. The defect Hamiltonian $H(\bf{k}, {\bf r})$ is described by a non-Hermitian Hamiltonian with spatially modulated adiabatical parameter ${\bf r}$ surrounding the defect. While the non-Hermitian system with point gap belongs to any of 38 symmetry classes (Bernard-LeClair classes), for non-Hermitian systems with line-like gap we get 54 non-equivalent generalized Bernard-LeClair classes as a natural generalization of  point gap classes. Although the classification of defects in Hermitian systems has been explored in the context of standard ten-fold Altland-Zirnbauer symmetry classes, a complete understanding of the role of the general non-Hermitian symmetries on the topological defects and their associated classification are still lacking.
           By continuous transformation and homeomorphic mapping, these non-Hermitian defect systems can be mapped to topologically equivalent Hermitian systems with associated symmetries,  and we get the topological classification by classifying the corresponding Hermitian Hamiltonians. We discuss some non-trivial classes with point gap according to our classification table, and give explicitly the topological invariants for these classes. We also study some lattice or continuous models and discuss the correspondence between the topological number and zero modes at the topological defect.
          \end{abstract}
          \maketitle
          \section{Introduction}
            Topological band theory has  achieved great success in the past decades \cite{Hassan,Qi,Chiu-RMP,Thouless,Haldane,Kane1,Kane2,Fu1,Fu2,ZhangSC} and topological classification according to standard ten-fold Altland-Zirnbauer (AZ) symmetry classes, defined by time-reversal symmetry, particle-hole symmetry and their combination, has been well established \cite{Chiu,Furusaki,AZ,Kitaev2,Ludwig1,Ludwig2,Ludwig3,Ludwig4,Stone}.
            In the scheme of AZ classification, a topological phase can be characterized by either $\mathbb{Z}$ or $\mathbb{Z}_2$ number. Two well-known examples are Haldane model and Kane-Mele model characterized by Chern number \cite{Thouless,Haldane} and $\mathbb{Z}_2$ number\cite{Kane1,Kane2}, respectively. In general, the bulk topological number has a correspondence with it's stable edge gapless modes.
            This correspondence leads to quantized Hall conductance and spin current \cite{Hassan,Qi}. The stable gapless modes also occur at the non-trivial topological defects, which have been studied in field theory \cite{JRD1,JRD2} and condensed matter systems \cite{SSH,Teo-PRL,Kitaev1,Read,Slager-PRB,Slager-PRB2015}. Systematic classifications of topological defects in insulators and superconductors have been carried out for AZ symmetry classes \cite{Teo}.

             Recent experimental studies of non-Hermitian properties in optic systems, electrical systems and open quantum systems  \cite{Ruter2010,Peng2014,Feng2014,Konotop2016,Xiao2017,Weimann2017,Menke,Klett}  have stimulated the development of non-Hermitian physics \cite{Bender1,Bender2,AUeda,Longhi}. Motivated by these progresses, the scope of topological phase of matter has also been extended to non-Hermitian systems \cite{Esaki2011,Hu2011,ZhuBG,Yuce,Hatano,Rudner,Xiong,Xiao,Weimann,ElG,TElee1,TElee2,Leykam,Parto,Gong,Das,Bergholtz,WangZhong1,Torres1,Yin,JiangHui,Lieu,Shen,WangZhong2,Song,Budich,Yamamoto}.              It has been unveiled that non-Hermitian systems exhibit many different properties from the Hermitian systems, e.g., biorthonormal eigenvectors \cite{Bergholtz}, the existence of exception points \cite{Rotter,Heiss,Hassan2017,Hu2017}, unusual bulk-edge correspondence \cite{Xiong,Bergholtz,WangZhong1,Torres1} and emergence of non-Hermitian skin effect in non-reciprocal systems  \cite{WangZhong1,Bergholtz,LangLJ,ZhaoYX2019,Lee2018,Edvardsson2019,Regnault,Slager2019}. Effects of non-Hermiticity on defect states are also studied for some specific non-Hermitian topological models \cite{LangLJ2018,Yuce-2018,YiWei,WuYJ}.
             Except the fundamental interest of understanding these differences,  novel topological properties of non-Hermitian systems may bring potential application in topological lasers and high-sensitive sensors.
             Besides extensive studies on various topological non-Hermitian models \cite{Yin,JiangHui,TElee2,Leykam,Lieu,Shen,WangZhong2,Song}, classification of non-Hermitian topological phases has been carried out by different groups \cite{Gong,LiuCH,Sato,Zhou}. In a recent work, Gong et. al went an important step for the classification of non-Hermitian phases with non-spatial symmetries \cite{Gong} by considering only time-reversal, pseudo particle-hole and  sublattice symmetries. After that, the topological classification of non-Hermitian systems in the AZ classes with an additional reflection symmetry was given by us \cite{LiuCH}. In general, non-Hermitian systems have more types of fundamental non-spatial symmetries than their Hermitian counterparts, i.e., there are totally $38$ different classes according to Bernard-LeClair (BL) symmetry classes \cite{Bernard}, which can be viewed as a natural non-Hermitian generalization of the ten AZ random matrix classes \cite{Lieu2,Bernard}.  Very recently, the full classifications for non-Hermitian topological band systems are obtained by Sato et.al \cite{Sato} and Zhou et.al\cite{Zhou}.

             So far, the topological classification of non-Hermitian systems focused on the band systems, classification of topological defects in these systems is still lacking. Aiming to fill this gap, in this work we make a full classification of topological defects for non-Hermitian systems. Due to the complex nature of spectra,  non-Hermitian systems have the point-like and line-like gaps, which corresponds to different complex-spectral-flattening procedure \cite{Sato}.
             For non-Hermitian systems with point-like gaps, the classification of topological defect can be carried out in the scheme of BL symmetry classes, which yield $38$ different classes with the help of the equivalent transformation $H\rightarrow iH$ \cite{Bernard,Zhou}. However, for non-Hermitian systems with line-like gaps, we notice that the transformation $H\rightarrow iH$
             can not be treated as  an equivalent transformation as a real (imaginary) gap is changed to  an imaginary (real) gap by such a transformation. Consequently, we construct 54 generalized Bernard-LeClair (GBL) classes for these systems. For both types of complex-energy gaps, we can transform the classification problem of non-Hermitian Hamiltonian to the classification problem of Hermitian Hamiltonian by continuous transformation and homeomorphic mapping. Then we classify the Hermitian Hamiltonian by using the K theory and Clifford algebra \cite{Chiu,Furusaki,Kitaev2}.
               And we also construct the topological number of topological defects for some BL classes with point gap. Then we study some continuous models and lattice models, and find that there is a correspondence between defect's topological number and gapless modes at the defect in some of these models.

The paper is organized as follows. In section II, we first give an introductory description of symmetry of non-Hermitian system in the AZ, BL and GBL classes, and also a brief introduction of different gap types. In section III, we complete topological classification of defects in non-Hermitian systems with point gap, real gap and imaginary gap, respectively, for all BL classes or GBL classes. In Section IV, we construct topological numbers of topological defects for some BL classes. In Section V,we study several example systems. A summary is given in the last section.

\section{Brief review of non-Hermitian symmetries and gap types}
Before presenting our approach of topological classification of defects in non-Hermitian systems, we first review briefly symmetries of non-Hermitian systems in the AZ and BL classes and the definitions of point gap, real and imaginary gaps, and then introduce the GBL classes. An introduction to these notations and definitions is necessary for our further classification of defects.

\subsection{Altland-Zirnbauer Class}
For Hermitian topological systems, the topological classification according time-reversal symmetry, particle-hole symmetry and chiral symmetry is called the AZ symmetry class, which leads to the standard 10-fold symmetry classification. It is known that AZ classes are not able to fully describe the internal symmetries of non-Hermitian systems, which belong to  more general BL classes \cite{Bernard,Zhou,Sato}. For a non-Hermitian Hamiltonian in momentum space, the time-reversal symmetry is defined by
\begin{equation}
    T=U_T \mathcal{K}, \qquad
    T H(-{\bf k})=  H({\bf k}) T, \qquad
    \end{equation}
where $\mathcal{K}$ is a complex conjugation operator and $U_K$ is a unitary matrix. Similarly, the particle-hole symmetry is defined by
    \begin{equation}
        A=U_A \mathcal{R}, \qquad
        A H(-{\bf k})= -H({\bf k})A, \qquad
    \end{equation}
where $\mathcal{R}$ is a transpose operator and $U_A$ is a unitary matrix \cite{Sato}. Given that $T$ is the time-reversal operator and $A$
the particle-hole operator, $\Gamma=TA$ is the chiral symmetry operator. Similar to the Hermitian case, AZ classes are constructed by $T$, $A$ and $\Gamma$ (see Table II).

\subsection{Bernard-LeClair Class}
 Bernard and LeClair made a full classification for non-Hermitian random matrix \cite{Bernard}, and it is the basic building block of non-Hermitian symmetries. In general, there are four types of symmetries in the non-Hermitian case denoted by K, Q, C and P, which are defined by
 \begin{align}
    H=kH^*k^{-1}&, ~~kk^*=\eta_k \mathbb{I}, &K \textrm{ sym.} \label{eq:syms1} \\
    H=qH^\dagger q^{-1}&, ~~q^2=\mathbb{I}, &Q \textrm{ sym.}\label{eq:syms2}\\
    H=\epsilon_c cH^Tc^{-1}&, ~~cc^*=\eta_c \mathbb{I},&C \textrm{ sym.}\label{eq:syms3}\\
    H=-pHp^{-1}&, ~~p^2=\mathbb{I}, &P \textrm{ sym.}\label{eq:syms4}
    \end{align}
where k, q, c and p are unitary matrices and $\eta_k,\epsilon_c,\eta_c= \pm 1$. The four unitary matrices satisfy:
      \begin{equation}
        c=\epsilon_{pc}pcp^T, \quad k=\epsilon_{pk}pkp^T, \quad c=\epsilon_{qc}qcq^T, \quad p=\epsilon_{pq}qpq^\dagger,
      \end{equation}
with $\epsilon_{pc}, \epsilon_{pk}, \epsilon_{qc}, \epsilon_{pq}=\pm 1$. We can construct 63 symmetry classes by these symmetries and the sign of $\eta_k,\epsilon_c,\eta_c,\epsilon_{pc},\epsilon_{pk},\epsilon_{qc}$ and $\epsilon_{pq}$.
For the point gap system, one can redefine $H \rightarrow iH$ as it does not change the properties of point gap. Consequently, $H=-kH^*k^{-1}$ and $H=-qH^\dagger q^{-1}$ transform to K and Q symmetries, respectively.
The 63 symmetry classes reduce to 38 topological classes (Non, P, Q, K1-2, C1-4, PQ1-2, PK1-3, PC1-4, QC1-8, PQC1-12) up to a redefinition of equivalence \cite{Zhou,Sato} (see Table III).
For line gap systems,  the transformation $H \rightarrow iH$ can not be taken as an equivalent  transformation, and then we need more than 38 classes beyond the BL classes. To understand this, we would like to introduce definition of gap before going to the discussion of the generalization of BL classes.
\begin{figure}[h] 
          \includegraphics[width=3.3in]{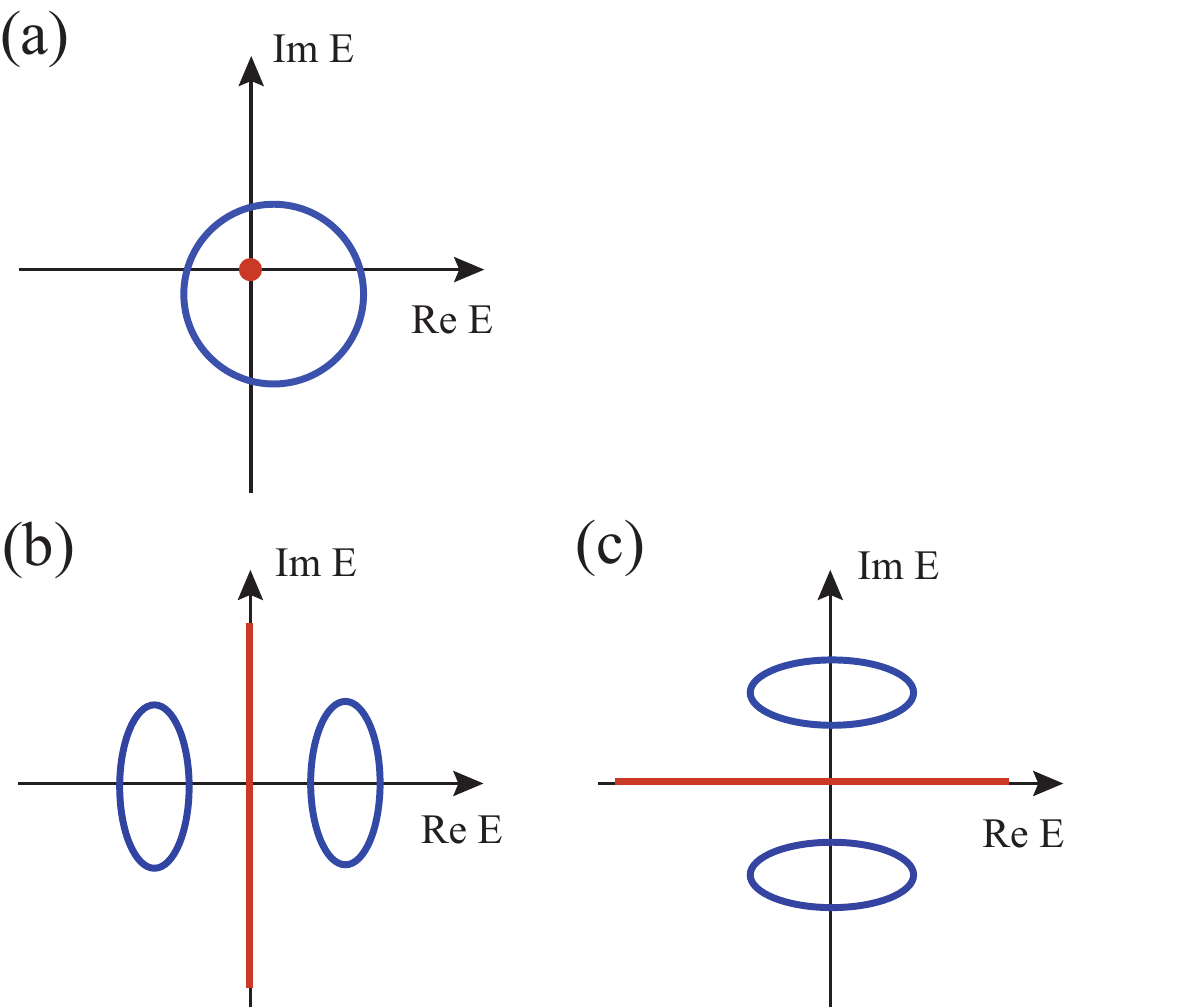}\\
          \caption{Schematic diagram of gap types. Red point (line) represents the gap, and blue areas represent the bands. (a) Point gap. (b) Real gap. (c) Imaginary gap.} \label{fig1}
         \end{figure}

\subsection{Point gap, real gap and imaginary gap }
In general, an energy gap in the band theory means a forbidden energy region with no occupancy of states. For the non-Hermitian systems, the definition of energy gap is nontrivial as the spectrum becomes complex. According to Kawabata et. al \cite{Sato}, non-Hermitian systems should have two different types of complex-energy gaps, i.e., the point-like and line-like gaps. Here we follow the definitions of Kawabata et. al.
Consider the complex energy plane, if a system has a point gap, it means that the band spectra can't cross the zero point (i.e., $\forall {\bf k}$, $det(H({\bf k})) \ne 0$) as schematically shown in Fig.(\ref{fig1}a).
        If a system has a real gap, it means that the band energies can't cross the imaginary axis (i.e., $ \forall j, {\bf k}$, $Re( E_j({\bf k})) \ne 0$) as shown in Fig.(\ref{fig1}b). If a system  has an imaginary gap, it means that the band energies can't cross the real axis (i.e., $ \forall j, {\bf k}$, $Im( E_j({\bf k}) )\ne 0 $) as shown in Fig.(\ref{fig1}c).
        Based on the three different constraints, we can get three different classifications.
        \begin{figure}[h] 
            \includegraphics[width=2.8in]{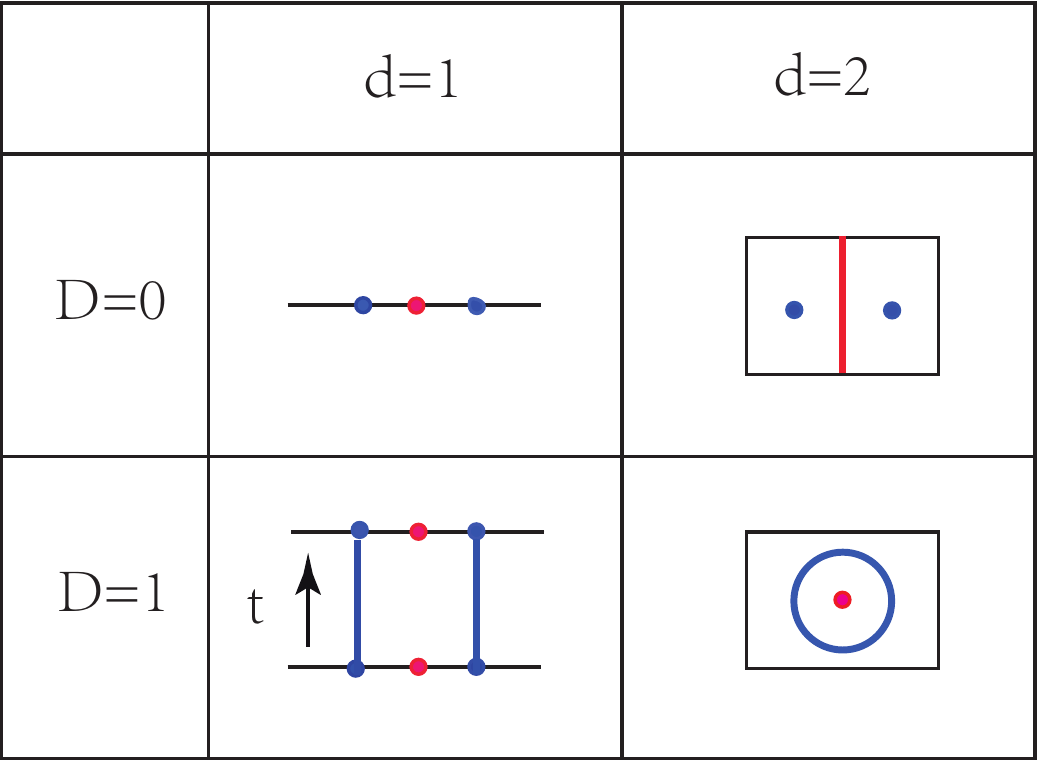}\\
            \caption{Schematic diagram of topological defects. A topological defect in $d$ dimension is surrounded by a $D$-dimensional surface $S^D$. For point defects, $d-D=1$. For line defects, $d-D=2$. } \label{fig2}
           \end{figure}

\subsection{Generalized Bernard-LeClair Class}
Since the transformation $H \rightarrow iH$  isn't an equivalent transformation for line gap systems, we now construct all nonequivalent non-Hermitian classes without considering $H \rightarrow iH$ as an equivalent transformation. In this case:
\begin{align}
    H=\epsilon_kkH^*k^{-1}&, ~~kk^*=\eta_k \mathbb{I}, &K \textrm{ sym.} \label{eq:syms1} \\
    H=\epsilon_qqH^\dagger q^{-1}&, ~~q^2=\mathbb{I}, &Q \textrm{ sym.}\label{eq:syms2}\\
    H=\epsilon_c cH^Tc^{-1}&, ~~cc^*=\eta_c \mathbb{I},&C \textrm{ sym.}\label{eq:syms3}\\
    H=-pHp^{-1}&, ~~p^2=\mathbb{I}, &P \textrm{ sym.}\label{eq:syms4}
    \end{align}
    were $\epsilon_k=\pm1$, $\epsilon_q=\pm1$ and other notations are the same as the BL class. We can construct 54 nonequivalent symmetry classes (Non, P, Qa-b, K1-2a-b, C1-4, PQ1-2, PK1-2, PK3a-b, PC1-4, QC1-8a-b, PQC1-8, PQC9-12a-b) by these symmetries and the sign of $\epsilon_k,\epsilon_q,\eta_k,\epsilon_c,\eta_c,\epsilon_{pc},\epsilon_{pk},\epsilon_{qc}$ and $\epsilon_{pq}$ and equivalence relations (see Table IV,V). The equivalence relations are similar to the equivalence relationships described by Zhou et al\cite{Zhou}, except that $\epsilon_{\tilde{k}}=-\epsilon_k$ and $\epsilon_{\tilde{q}}=-\epsilon_q$ when we consider P and K symmetries or P and Q symmetries ($\tilde{k}=pk$ and $\tilde{q}=\sqrt{\epsilon_{pq}}qp$) \cite{Zhou}.
   To distinguish with the previous BL class which supports 38 different classes, we call this class as GBL class.

\section{Topological classification }
A non-Hermitian periodic system with topological defects is described by $H({\bf k}, {\bf r})$, where ${\bf k}$ is defined in a d-dimensional Brillouin zone $T^d$, and ${\bf r}$ is defined on a D-dimensional surface $S^D$ surrounding the defect (Fig.\ref{fig2}). The defect Hamiltonian is a band Hamiltonian slowly modulated by a parameter r, which
includes spatial coordinates and/or a temporal parameter and changes slow enough so that the bulk system separated far
from the defect core still can be characterized by momentum ${\bf k}$ \cite{Teo} .

\subsection{ Point gap classification }
If the system has a point gap, we can map the non-Hermitian Hamiltonian to a Hermitian one by the doubling process \cite{Gong,RH,Sato,Bernard}:
          \begin{equation}
            \tilde {H}({\bf k}, {\bf r})=\left[ \begin{array}{cc}
                0 & H({\bf k}, {\bf r})\\
                H({\bf k}, {\bf r})^\dagger & 0
                \end{array}
                \right ]. \label{H}
            \end{equation}
The doubled Hamiltonian $\tilde {H}({\bf k}, {\bf r})$ fulfills an enforced additional chiral symmetry: 
              \begin{equation}
                \Sigma \tilde {H}({\bf k}, {\bf r})=  -\tilde {H}({\bf k}, {\bf r}) \Sigma , \label{chiralsym}
                \end{equation}
where $\Sigma = \sigma_z \otimes 1$ and $\Sigma^2=1$. And the constraint on $det(H({\bf k}, {\bf r})) \ne 0$ is equivalent to $det(\tilde {H}({\bf k}, {\bf r})) \ne 0 $. Then $H({\bf k}, {\bf r})$ is homeomorphic to $\tilde {H}({\bf k}, {\bf r})$, and it is equivalent to classify Hermitian Hamiltonian $\tilde {H}({\bf k}, {\bf r})$ with chiral symmetry. If
$H({\bf k}, {\bf r})$ has $T$, $A$, $\Gamma$, $K$, $Q$, $P$ and $C$ symmetries,  $\tilde {H}({\bf k}, {\bf r})$ has the corresponding symmetries:
 \begin{equation}
     \tilde{T} \tilde {H}({\bf k}, {\bf r})= \tilde {H}(-{\bf k}, {\bf r})\tilde{T},
 \end{equation}
 \begin{equation}
    \tilde{A} \tilde {H}({\bf k}, {\bf r})= - \tilde {H}(-{\bf k}, {\bf r})\tilde{A},
\end{equation}
\begin{equation}
    \tilde{\Gamma} \tilde {H}({\bf k}, {\bf r})= - \tilde {H}({\bf k}, {\bf r})\tilde{\Gamma},
\end{equation}
\begin{equation}
    \tilde{K} \tilde {H}({\bf k}, {\bf r})=\tilde {H}(-{\bf k}, {\bf r})\tilde{K},
\end{equation}
\begin{equation}
    \tilde{Q} \tilde {H}({\bf k}, {\bf r})=\tilde {H}({\bf k}, {\bf r})\tilde{Q},
\end{equation}
\begin{equation}
    \tilde{P} \tilde {H}({\bf k}, {\bf r})=-\tilde {H}({\bf k}, {\bf r})\tilde{P},
\end{equation}
\begin{equation}
    \tilde{C} \tilde {H}({\bf k}, {\bf r})=\epsilon_c \tilde {H}(-{\bf k}, {\bf r})\tilde{C},  \label{c}
\end{equation}
  with $\tilde{T}=\sigma_0\otimes U_T\mathcal{K}$, $\tilde{A}=\sigma_x\otimes U_A\mathcal{K}$, $\tilde{\Gamma}=\tilde{T}\tilde{A}$, $\tilde{K}=\sigma_0 \otimes k\mathcal{K}$, $\tilde{Q}=\sigma_x\otimes q$, $\tilde{P}=\sigma_0 \otimes p$ and $\tilde{C}=\sigma_x\otimes c\mathcal{K}$.

  Generally, we can represent the defect Hamiltonian as:
      \begin{equation}
        \tilde {H}({\bf k}, {\bf r})=\tilde{\gamma}_0+ k_1\tilde{\gamma}_1^k+...+k_d \tilde{\gamma}_d^k +r_1\tilde{\gamma}_1^r+...+r_D\tilde{\gamma}_D^r ,
     \end{equation}
where
        $\tilde{\gamma}_0,\tilde{\gamma}_i^k(i=1,...,d)$ and $\tilde{\gamma}_j^r(j=1,...,D)$ anticommute with each other and their squares equal to the identity operator. Using the commutation relations of symmetry operators and Hamiltonian, we can construct the  Clifford algebra's extension for each symmetry class. Then we can get the space of the mass term ($\tilde{\gamma_0}$) by the correspondence between the Clifford algebra's extension and space of the mass term (Table I).
\begin{table}[h]
            \caption{\label{tab:table1} The correspondence between the Clifford algebra's extension and space of the mass term.}
          \begin{ruledtabular}
          \begin{tabular}{cc}
            Clifford algebra's extension & Space of the mass term\\
            \hline
            $Cl_i \rightarrow Cl_{i+1}$&$C_i=C_{i+2}$\\
            $Cl_{p,q} \rightarrow Cl_{p,q+1}$&$R_{0,q-p}=R_{0,q-p+8}$\\
            $Cl_{p,q} \rightarrow Cl_{p+1,q}$&$R_{0,p-q+2}=R_{0,p-q+10}$\\
            \end{tabular}
          \end{ruledtabular}
        \end{table}

        For convenience, here we briefly introduce Clifford algebra. There are two types of Clifford algebra, i.e., the complex Clifford algebra and real Clifford algebra. Complex Clifford algebra is defined in the complex domain, and its generators $\{\gamma_1,...,\gamma_n\}$ satisfy that $\{\gamma_i,\gamma_j\}=2\delta_{ij}$. Complex Clifford algebra can be represented as $Cl_n$. On the other hand, real Clifford algebra is defined in the real domain, and its generators $\{\gamma_1^-,...,\gamma_{n1}^-,\gamma_1^+,...,\gamma_{n2}^+\}$  anticommute with each others. While $\gamma_i^-(i=1,...,n1)$ are squared to $-1$, $\gamma_j^+(j=1,...,n2)$ are squared to $1$. The real Clifford algebra can be represented as $Cl_{n1,n2}$.

\begin{table*}[htbp]
    \begin{center}
        \begin{ruledtabular}
    \begin{tabular}{c|ccc|ccccccccc}
    \multicolumn{4}{c|}{Symmetry } & \multicolumn{9}{c}{  $\delta = d - D$} \\
     \multicolumn{1}{c}{AZ} &$U_TU_T^*$ &$U_AU_A^*$ & $ \Gamma^2$ &Cl&$0$&
    $1$   &  $2$ &  $3$ &  $4$ &  $5$ & $6$ & $7$\\
     \hline
     A (Non) & $0$ & $0$ & $0$ &$C_1$&
     $0$ & $\mathbb{Z}$ & $0$ & $\mathbb{Z}$ & $0$ & $\mathbb{Z}$ & $0$ & $\mathbb{Z}$\\
     AIII (Q) & $0$ & $0$ & $1$ &$C_0$&
     $\mathbb{Z}$ & $0$ & $\mathbb{Z}$ & $0$ & $\mathbb{Z}$ & $0$ & $\mathbb{Z}$ & $0$\\
    \hline
     AI (K1)&  $1$ & $0$ & $0$ &$R_1$&
           $\mathbb{Z}_2$ & $\mathbb{Z}$ & $0$ & $0$ & $0$ & $2\mathbb{Z}$ & $0$ & $\mathbb{Z}_2$\\
     BDI (QC5) & $1$ & $1$ &$1$ &$R_2$&
          $\mathbb{Z}_2$ & $\mathbb{Z}_2$ & $\mathbb{Z}$ & $0$ & $0$ & $0$ & $2\mathbb{Z}$ & $0$\\
     D (C3) & $0$ & $1$ & $0$ &$R_3$&
          $0$ & $\mathbb{Z}_2$ & $\mathbb{Z}_2$ & $\mathbb{Z}$ & $0$ & $0$ & $0$& $2\mathbb{Z}$ \\
    DIII (QC6) & $-1$ & $1$ & $1$ &$R_4$&
         $2\mathbb{Z}$ & $0$ & $\mathbb{Z}_2$ & $\mathbb{Z}_2$ & $\mathbb{Z}$ & $0$ & $0$ & $0$\\
    AII (K2) & $-1$ & $0$ & $0$ &$R_5$&
         $0$ & $2\mathbb{Z}$ & $0$ & $\mathbb{Z}_2$ & $\mathbb{Z}_2$ & $\mathbb{Z}$ & $0$ & $0$\\
     CII (QC7) & $-1$ & $-1$ & $1$ &$R_6$&
          $0$ & $0$ & $2\mathbb{Z}$ & $0$ & $\mathbb{Z}_2$ & $\mathbb{Z}_2$ & $\mathbb{Z}$ & $0$\\
     C (C4) &  $0$ & $-1$ & $0$ &$R_7$&
          $0$ & $0$ & $0$ & $2\mathbb{Z}$ & $0$ & $\mathbb{Z}_2$ & $\mathbb{Z}_2$ & $\mathbb{Z}$\\
     CI (QC8) &  $1$ & $-1$ & $1$ &$R_0$&
          $\mathbb{Z}$ & $0$ & $0$ & $0$ & $2\mathbb{Z}$ & $0$ & $\mathbb{Z}_2$ & $\mathbb{Z}_2$\\
    \end{tabular}

    \caption{Periodic table for point gap classification of topological defects in non-Hermitian systems.  The rows correspond to the different AZ symmetry
    classes, while the columns depend
    on $\delta = d-D$.}
    \end{ruledtabular}
    \end{center}
    \label{tab:AZ periodic}
    \end{table*}

Once we know the space of the mass term, we can get the topological classification by calculating the zero-order homotopy group of the space of the mass term. For the case of point gap,
we give the classification according to the symmetry classes labeled by both standard AZ classes and  BL classes, despite that the former ones are the subclasses of the latter ones. For convenience, when we discuss the AZ classes, we also label the corresponding BL classes simultaneously.
Firstly, we consider the complex classes:

        {\bf Class A (Non)}: The generators of this class are $\left\{\tilde{\gamma}_0,\tilde{\gamma}_1^k,...,\tilde{\gamma}_d^k,\tilde{\gamma}_1^r,...,\tilde{\gamma}_D^r,\Sigma\right\}$ (The generators of class are different from generators of Clifford algebra). The Clifford algebra's extension of this class is $\left\{\tilde{\gamma}_1^k,...,\tilde{\gamma}_d^k,\tilde{\gamma}_1^r,...,\tilde{\gamma}_D^r,\Sigma\right\} \rightarrow \left\{\tilde{\gamma}_0,\tilde{\gamma}_1,...,\tilde{\gamma}_d^k,\tilde{\gamma}_1^r,...,\tilde{\gamma}_D^r,\Sigma\right\}=Cl_{d+D+1}\rightarrow Cl_{d+D+2}=Cl_{1-(d-D)}\rightarrow Cl_{2-(d-D)}$. The space of the mass term ($\tilde{\gamma_0}$) is $C_{1-\delta}$ $(\delta=d-D)$, and the topological classification of defects for the non-Hermitian class A is characterized by $\pi_0(C_{1-\delta})=0 ~(\mathbb{Z})$ for  even (odd) $\delta$. And the classifying space (CL) is $C_1$ (classifying space is equivalent to the space of mass term when $\delta=0$).

        {\bf Class AIII (Q)}: The generators of this class are $\{\tilde{\gamma}_0,\tilde{\gamma}_1^k,...,\tilde{\gamma}_d^k,\tilde{\gamma}_1^r,...,\tilde{\gamma}_D^r,\Sigma,\tilde{\Gamma} \}$. The Clifford algebra's extension of this class is $\{\tilde{\gamma}_1^k,...,\tilde{\gamma}_d^k,\tilde{\gamma}_1^r,...,\tilde{\gamma}_D^r,\Sigma,\tilde{\Gamma}\} \rightarrow \{\tilde{\gamma}_0,\tilde{\gamma}_1^k,...,\tilde{\gamma}_d^k,\tilde{\gamma}_1^r,...,\tilde{\gamma}_D^r,\Sigma,\tilde{\Gamma}\}=Cl_{d+D+2}\rightarrow Cl_{d+D+3}$$=Cl_{2-(d-D)}\rightarrow Cl_{3-(d-D)}$. The space of the mass term ($\tilde{\gamma_0}$) is $C_{2-\delta}$ $(\delta=d-D)$, and the classification is characterized by $\pi_0(C_{2-\delta})=\mathbb{Z}(0)$ for  even (odd) $\delta$.

        Then we classify the real classes:

        {\bf Class AI (K1) }: The generators of this class are $\{\tilde{\gamma}_0,\tilde{\gamma}_1^k,...,\tilde{\gamma}_d^k,\tilde{\gamma}_1^r,...,\tilde{\gamma}_D^r,\Sigma ,\tilde{T}, J \}$, where $J$ is the imaginary unit. The Clifford algebra's extension of this class is $\{\tilde{\gamma}_1^k,...,\tilde{\gamma}_d^k,J\tilde{\gamma}_1^r,...,J\tilde{\gamma}_D^r,J\Sigma,\tilde{T},J\tilde{T} \} \rightarrow \{ J\tilde{\gamma}_0,\tilde{\gamma}_1^k,...,\tilde{\gamma}_d^k,J\tilde{\gamma}_1^r,...,J\tilde{\gamma}_D^r,J\Sigma,\tilde{T}, J\tilde{T} \}=Cl_{D+1,d+2}\rightarrow Cl_{D+2,d+2}$. The space of the mass term is $R_{1-\delta}$.

        {\bf Class C (C4)}: The generators of this class are $\{\tilde{\gamma}_0,\tilde{\gamma}_1^k,...,\tilde{\gamma}_d^k,\tilde{\gamma}_1^r,...,\tilde{\gamma}_D^r,\Sigma ,\tilde{A}, J \}$. The Clifford aglebra's extension of this class is $\{J\tilde{\gamma}_1^k,...,J\tilde{\gamma}_d^k,\tilde{\gamma}_1^r,...,\tilde{\gamma}_D^r,\Sigma,\tilde{A},J\tilde{A} \} \rightarrow \{ \tilde{\gamma}_0,J\tilde{\gamma}_1^k,...,J\tilde{\gamma}_d^k,\tilde{\gamma}_1^r,...,\tilde{\gamma}_D^r,\Sigma,\tilde{A}, J\tilde{A} \}=Cl_{d+2,D+1}\rightarrow Cl_{d+2,D+2}$. The space of the mass term is $R_{7-\delta}$.

        {\bf Class BDI (QC5)}: The generators of this class are $\{\tilde{\gamma}_0,\tilde{\gamma}_1^k,...,\tilde{\gamma}_d^k,\tilde{\gamma}_1^r,...,\tilde{\gamma}_D^r,\Sigma ,\tilde{A} ,\tilde{\Gamma} , J \}$. The Clifford algebra's extension of this class is $\{J\tilde{\gamma}_1^k,...,J\tilde{\gamma}_d^k,\tilde{\gamma}_1^r,...,\tilde{\gamma}_D^r,\Sigma, J\tilde{\Gamma}, \tilde{A}, J\tilde{A} \} \rightarrow \{ \tilde{\gamma}_0,J\tilde{\gamma}_1^k,...,J\tilde{\gamma}_d^k,\tilde{\gamma}_1^r,...,\tilde{\gamma}_D^r,\Sigma, J\tilde{\Gamma}, \tilde{A}, J\tilde{A} \}=Cl_{d+1,D+3}\rightarrow Cl_{d+1,D+4}$. The space of the mass term is $R_{2-\delta}$.

Similarly, we can determine the space of mass term for the other AZ classes.
It was shown that the classification of topological defects depends only on the topological dimension $\delta=d-D$. Once the classifying space is known, e.g. $R_{n}$, we can get the space of mass term in a given topological dimension by a shift of $R_{n}$ to $R_{n-\delta}$, and the topological classification is obtained by calculating $\pi_0 (R_{n-\delta})$. The topological classification of defects for all AZ classes are summarized in table II.

In the same way, we can get the topological classification of BL classes. For example,
for the  class {\bf P}, the generators are $\{\tilde{\gamma}_0,\tilde{\gamma}_1^k,...,\tilde{\gamma}_d^k,\tilde{\gamma}_1^r,...,\tilde{\gamma}_D^r,\Sigma ,\tilde{P} \}$, and the Clifford algebra's extension of this class is $\{\tilde{\gamma}_1^k,...,\tilde{\gamma}_d^k,\tilde{\gamma}_1^r,...,\tilde{\gamma}_D^r,\Sigma \} \otimes \{ \Sigma\tilde{P} \} \rightarrow \{ \tilde{\gamma}_0, \tilde{\gamma}_1^k,..., \tilde{\gamma}_d^k, \tilde{\gamma}_1^r,...,\tilde{\gamma}_D^r,\Sigma \} \otimes \{ \Sigma\tilde{P} \}=Cl_{d+D+1}\times Cl_{d+D+1}\rightarrow Cl_{d+D+2}\times Cl_{d+D+2}$. It follows that the space of the mass term is $C_{1-\delta}\times C_{1-\delta}$.
The space of mass term for other BL classes can be obtained under the same scheme. The results are listed in table III.  For classes Cn and C5-n $(n=1,\cdots,4)$, the Hermitianized Hamiltonians fulfill the same symmetry constraints, i.e., Eq.(\ref{chiralsym}) and Eq.(\ref{c}). Then classes Cn and C5-n have the same topological classification. Similarly, classes QCn and QC9-n $(n=1,\cdots,8)$ have the same topological classification. When $D=0$, we have $\delta=d$ and our results can be applied
to describe the classification of point gap systems in the absence of defect, which are consistent with those in the reference \cite{Sato,Zhou}. For the convenience of comparing with results in references \cite{Zhou,Sato}, we list the correspondence between different notations of BL class in the table VII of the appendix C.

    \subsection{Real gap classification}
    If a non-Hermitian Hamiltonian has a real gap, it has been demonstrated that the Hamiltonian can continuously transform to a Hermitian Hamiltonian $H$ while keeping its symmetry and real gap \cite{Sato}. Then the non-Hermitian GBL class classification is the same with the corresponding Hermitian  classification. To classify 54 GBL classes, the K, Q, C and P symmetries reduce to:
    \begin{align}
      H=\epsilon_k kH^*k^{-1}&, kk^*=\eta_k \mathbb{I}&K \textrm{ sym.} \label{eq:syms1} \\
      H=\epsilon_q qH q^{-1}&, q^2=\mathbb{I}&Q \textrm{ sym.}\label{eq:syms2}\\
      H=\epsilon_c cH^* c^{-1}&, cc^*=\eta_c \mathbb{I}&C \textrm{ sym.}\label{eq:syms3}\\
      H=-pHp^{-1}&, p^2=\mathbb{I}&P \textrm{ sym.}\label{eq:syms4}
      \end{align}
    Here $H$ is a Hermitian Hamiltonian. The classification problem reduces to a Hermitian Hamiltonian classification problem. And we can classify each class by Clifford algebra (See Appendix A). The results are listed in Table IV.  When $D=0$, our results can be applied to describe the classification of real gap systems in the absence of defect. In order to compare with results in the reference \cite{Sato}, we list the correspondence between our notations and those in the reference \cite{Sato} in the table VIII of the appendix C.

    \subsection{Imaginary gap classification}
    If a non-Hermitian Hamiltonian has an imaginary gap, the Hamiltonian can continuously transform to an anti-Hermitian Hamiltonian $iH$ \cite{Sato}, where $H$ is a Hermitian Hamiltonian. To classify 54 GBL classes, the K, Q, C and P symmetries reduce to:
    \begin{align}
      H=-\epsilon_k kH^*k^{-1}&, kk^*=\eta_k \mathbb{I}&K \textrm{ sym.} \label{eq:syms1} \\
      H=-\epsilon_q qH q^{-1}&, q^2=\mathbb{I}&Q \textrm{ sym.}\label{eq:syms2}\\
      H=\epsilon_c cH^*c^{-1}&, cc^*=\eta_c \mathbb{I}&C \textrm{ sym.}\label{eq:syms3}\\
      H=-pHp^{-1}&, p^2=\mathbb{I}&P \textrm{ sym.}\label{eq:syms4}
      \end{align}
    The classification problem reduces to a Hermitian Hamiltonian classification problem. And we can classify each class by Clifford algebra (See Appendix B). The results are listed in Table V. Similarly, our results can be applied to describe the classification of imaginary gap systems in the absence of defect by taking $D=0$. The correspondence between our notations and those in the reference \cite{Sato} can be found in the table VIII of the appendix C.

    \onecolumngrid \clearpage
    \begin{table*}[htbp]\footnotesize
      \begin{center}
      \caption{\label{tab:table4}  Periodic table of point gap classification of topological defects in non-Hermitian systems.  The rows correspond to the different BL symmetry
      classes, while the columns depend
      on $\delta = d-D$.  The topological numbers in the table are stable strong topological numbers.}
    \begin{tabular}{|c|c|c|cccccccc|}
        \hline
        BL&Gen. Rel.&Cl&$\delta=0$&1&2\;\;\;&\;\;\;3\;\;\;&\;\;\;4\;\;\;&\;\;\;\;5\;\;\;\;&\;\;\;\;6\;\;\;\;&\;\;\;\;7\;\;\;\\
      \hline
    Non&  &$C_1$&$0$ & $\mathbb{Z}$ & $0$ & $\mathbb{Z}$ & $0$ & $\mathbb{Z}$ & $0$ & $\mathbb{Z}$\\
      \hline
    P&   &$C_1 ^2 $& $0\oplus 0$ & $\mathbb{Z} \oplus \mathbb{Z}$ & $0\oplus 0$ & $\mathbb{Z} \oplus \mathbb{Z}$ & $0\oplus 0$ & $\mathbb{Z} \oplus \mathbb{Z}$ & $0\oplus 0$ & $\mathbb{Z} \oplus \mathbb{Z}$\\
      \hline
    Q&  & $ C_0 $&$\mathbb{Z}$ & $0$ & $\mathbb{Z}$ & $0$ & $\mathbb{Z}$ & $0$ & $\mathbb{Z}$ & $0$\\
      \hline
    K1& $\eta_k =1$ & $ R_1 $ & $\mathbb{Z}_2$ & $\mathbb{Z}$ & $0$ & $0$ & $0$ & $2\mathbb{Z}$ & $0$ & $\mathbb{Z}_2$\\
      \hline
    K2& $\eta_k =-1$ & $ R_5 $& $0$ & $2\mathbb{Z}$ & $0$ & $\mathbb{Z}_2$ & $\mathbb{Z}_2$ & $\mathbb{Z}$ & $0$ & $0$\\
      \hline
    C1& $\epsilon_c=1$, $\eta_c=1$ & $ R_7 $& $0$ & $0$ & $0$ & $2\mathbb{Z}$ & $0$ & $\mathbb{Z}_2$ & $\mathbb{Z}_2$ & $\mathbb{Z}$\\
      \hline
    C2& $\epsilon_c=1$, $\eta_c=-1$ & $ R_3 $& $0$ & $\mathbb{Z}_2$ & $\mathbb{Z}_2$ & $\mathbb{Z}$ & $0$ & $0$ & $0$& $2\mathbb{Z}$ \\
      \hline
    C3& $\epsilon_c=-1$, $\eta_c=1$ & $ R_3 $& $0$ & $\mathbb{Z}_2$ & $\mathbb{Z}_2$ & $\mathbb{Z}$ & $0$ & $0$ & $0$& $2\mathbb{Z}$ \\
      \hline
    C4& $\epsilon_c=-1$, $\eta_c=-1$ & $ R_7 $& $0$ & $0$ & $0$ & $2\mathbb{Z}$ & $0$ & $\mathbb{Z}_2$ & $\mathbb{Z}_2$ & $\mathbb{Z}$\\
      \hline
    PQ1& $\epsilon_{pq}=1$ & $ C_1 $& $0$ & $\mathbb{Z}$ & $0$ & $\mathbb{Z}$ & $0$ & $\mathbb{Z}$ & $0$ & $\mathbb{Z}$\\
    \hline
    PQ2& $\epsilon_{pq}=-1$ & $ C_0 ^2 $& $\mathbb{Z} \oplus \mathbb{Z}$ & $0\oplus 0$ & $\mathbb{Z} \oplus \mathbb{Z}$ & $0\oplus 0$ & $\mathbb{Z} \oplus \mathbb{Z}$ & $0\oplus 0$ & $\mathbb{Z} \oplus \mathbb{Z}$ &  $0\oplus 0$ \\
    \hline
    PK1& $\eta_k=1$, $\epsilon_{pk}=1$ & $ R_1 ^2 $& $\mathbb{Z}_2 \oplus \mathbb{Z}_2$ & $\mathbb{Z} \oplus \mathbb{Z}$ & $0 \oplus 0$ & $0 \oplus 0$ & $0 \oplus 0$ & $2\mathbb{Z}\oplus 2\mathbb{Z}$ & $0 \oplus 0$ & $\mathbb{Z}_2 \oplus \mathbb{Z}_2$\\
    \hline
    PK2& $\eta_k=-1$, $\epsilon_{pk}=1$ & $ R_5 ^2 $& $0\oplus 0$ & $2\mathbb{Z}\oplus 2\mathbb{Z}$ & $0\oplus 0$ & $\mathbb{Z}_2\oplus \mathbb{Z}_2$ & $\mathbb{Z}_2 \oplus \mathbb{Z}_2$ & $\mathbb{Z}\oplus \mathbb{Z}$ & $0\oplus 0$ & $0\oplus 0$\\
    \hline
   PK3& \begin{tabular}{c} $\eta_k=1$, $\epsilon_{pk}=-1$ \\ \hline $\eta_k=-1$, $\epsilon_{pk}=-1$    \end{tabular} & $ C_1 $& $0$ & $\mathbb{Z}$ & $0$ & $\mathbb{Z}$ & $0$ & $\mathbb{Z}$ & $0$ & $\mathbb{Z}$\\
      \hline
    PC1& \begin{tabular}{c} $\epsilon_c=1$, $\eta_c=1$, $\epsilon_{pc}=1$ \\ \hline $\epsilon_c=-1$, $\eta_c=1$, $\epsilon_{pc}=1$ \end{tabular} & $ C_1 $& $0$ & $\mathbb{Z}$ & $0$ & $\mathbb{Z}$ & $0$ & $\mathbb{Z}$ & $0$ & $\mathbb{Z}$\\
      \hline
    PC2& \begin{tabular}{c} $\epsilon_c=1$, $\eta_c=1$, $\epsilon_{pc}=-1$ \\ \hline $\epsilon_c=-1$, $\eta_c=-1$, $\epsilon_{pc}=-1$ \end{tabular} & $ R_7 ^2 $& $0\oplus 0$ & $0\oplus 0$ & $0\oplus 0$ & $2\mathbb{Z}\oplus 2\mathbb{Z}$ & $0\oplus 0$ & $\mathbb{Z}_2\oplus \mathbb{Z}_2$ &  $\mathbb{Z}_2\oplus \mathbb{Z}_2$ & $\mathbb{Z}\oplus \mathbb{Z}$\\
      \hline
    PC3& \begin{tabular}{c} $\epsilon_c=1$, $\eta_c=-1$, $\epsilon_{pc}=1$ \\ \hline $\epsilon_c=-1$, $\eta_c=-1$, $\epsilon_{pc}=1$ \end{tabular} & $ C_1 $& $0$ & $\mathbb{Z}$ & $0$ & $\mathbb{Z}$ & $0$ & $\mathbb{Z}$ & $0$ & $\mathbb{Z}$\\
      \hline
    PC4& \begin{tabular}{c} $\epsilon_c=1$, $\eta_c=-1$, $\epsilon_{pc}=-1$ \\ \hline $\epsilon_c=-1$, $\eta_c=1$, $\epsilon_{pc}=-1$ \end{tabular} & $ R_3 ^2 $&  $0\oplus 0$ & $\mathbb{Z}_2\oplus \mathbb{Z}_2$ & $\mathbb{Z}_2\oplus \mathbb{Z}_2$ & $\mathbb{Z}\oplus \mathbb{Z}$ & $0\oplus 0$ & $0\oplus 0$ & $0\oplus 0$& $2\mathbb{Z}\oplus 2\mathbb{Z}$ \\
      \hline
    QC1& $\epsilon_c=1$, $\eta_c=1$, $\epsilon_{qc}=1$ & $ R_0 $&  $\mathbb{Z}$ & $0$ & $0$ & $0$ & $2\mathbb{Z}$ & $0$ & $\mathbb{Z}_2$ & $\mathbb{Z}_2$\\
      \hline
    QC2&  $\epsilon_c=1$, $\eta_c=1$, $\epsilon_{qc}=-1$  & $ R_6 $& $0$ & $0$ & $2\mathbb{Z}$ & $0$ & $\mathbb{Z}_2$ & $\mathbb{Z}_2$ & $\mathbb{Z}$ & $0$\\
      \hline
    QC3&  $\epsilon_c=1$, $\eta_c=-1$, $\epsilon_{qc}=1$  & $ R_4 $& $2\mathbb{Z}$ & $0$ & $\mathbb{Z}_2$ & $\mathbb{Z}_2$ & $\mathbb{Z}$ & $0$ & $0$ & $0$\\
      \hline
    QC4& $\epsilon_c=1$, $\eta_c=-1$, $\epsilon_{qc}=-1$  & $ R_2 $&  $\mathbb{Z}_2$ & $\mathbb{Z}_2$ & $\mathbb{Z}$ & $0$ & $0$ & $0$ & $2\mathbb{Z}$ & $0$\\
      \hline
    QC5& $\epsilon_c=-1$, $\eta_c=1$, $\epsilon_{qc}=1$  & $ R_2 $&  $\mathbb{Z}_2$ & $\mathbb{Z}_2$ & $\mathbb{Z}$ & $0$ & $0$ & $0$ & $2\mathbb{Z}$ & $0$\\
      \hline
    QC6& $\epsilon_c=-1$, $\eta_c=1$, $\epsilon_{qc}=-1$  & $ R_4 $& $2\mathbb{Z}$ & $0$ & $\mathbb{Z}_2$ & $\mathbb{Z}_2$ & $\mathbb{Z}$ & $0$ & $0$ & $0$\\
      \hline
    QC7& $\epsilon_c=-1$, $\eta_c=-1$, $\epsilon_{qc}=1$  & $ R_6 $& $0$ & $0$ & $2\mathbb{Z}$ & $0$ & $\mathbb{Z}_2$ & $\mathbb{Z}_2$ & $\mathbb{Z}$ & $0$\\
      \hline
    QC8& $\epsilon_c=-1$, $\eta_c=-1$, $\epsilon_{qc}=-1$   & $ R_0 $&  $\mathbb{Z}$ & $0$ & $0$ & $0$ & $2\mathbb{Z}$ & $0$ & $\mathbb{Z}_2$ & $\mathbb{Z}_2$\\
      \hline
    PQC1& \begin{tabular}{c} $\epsilon_c=1$, $\eta_c=1$, $\epsilon_{pq}=1$, $\epsilon_{pc}=1$, $\epsilon_{qc}=1$ \\ \hline $\epsilon_c=-1$, $\eta_c=1$, $\epsilon_{pq}=1$, $\epsilon_{pc}=1$, $\epsilon_{qc}=1$ \end{tabular} & $ R_1 $&   $\mathbb{Z}_2$ & $\mathbb{Z}$ & $0$ & $0$ & $0$ & $2\mathbb{Z}$ & $0$ & $\mathbb{Z}_2$\\
      \hline
    PQC2& \begin{tabular}{c} $\epsilon_c=1$, $\eta_c=1$, $\epsilon_{pq}=1$, $\epsilon_{pc}=1$, $\epsilon_{qc}=-1$ \\ \hline $\epsilon_c=-1$, $\eta_c=1$, $\epsilon_{pq}=1$, $\epsilon_{pc}=1$, $\epsilon_{qc}=-1$ \end{tabular} & $ R_5 $&  $0$ & $2\mathbb{Z}$ & $0$ & $\mathbb{Z}_2$ & $\mathbb{Z}_2$ & $\mathbb{Z}$ & $0$ & $0$\\
      \hline
    PQC3& \begin{tabular}{c} $\epsilon_c=1$, $\eta_c=1$, $\epsilon_{pq}=-1$, $\epsilon_{pc}=-1$, $\epsilon_{qc}=1$ \\ \hline $\epsilon_c=-1$, $\eta_c=-1$, $\epsilon_{pq}=-1$, $\epsilon_{pc}=-1$, $\epsilon_{qc}=-1$ \end{tabular} & $ R_0^2 $& $\mathbb{Z}\oplus \mathbb{Z}$ & $0\oplus 0$ & $0\oplus 0$ & $0\oplus 0$ & $2\mathbb{Z}\oplus 2\mathbb{Z}$ & $0\oplus 0$ & $\mathbb{Z}_2 \oplus \mathbb{Z}_2$ & $\mathbb{Z}_2 \oplus \mathbb{Z}_2$ \\
      \hline
    PQC4& \begin{tabular}{c} $\epsilon_c=1$, $\eta_c=1$, $\epsilon_{pq}=-1$, $\epsilon_{pc}=-1$, $\epsilon_{qc}=-1$ \\ \hline $\epsilon_c=-1$, $\eta_c=-1$, $\epsilon_{pq}=-1$, $\epsilon_{pc}=-1$, $\epsilon_{qc}=1$ \end{tabular} & $ R_6 ^2 $& $0\oplus 0$ & $0\oplus 0$ & $2\mathbb{Z}\oplus 2\mathbb{Z}$ & $0\oplus 0$ & $\mathbb{Z}_2 \oplus \mathbb{Z}_2$ &$\mathbb{Z}_2 \oplus \mathbb{Z}_2$ & $\mathbb{Z} \oplus \mathbb{Z}$ & $0\oplus 0$\\
      \hline
    PQC5& \begin{tabular}{c} $\epsilon_c=1$, $\eta_c=-1$, $\epsilon_{pq}=1$, $\epsilon_{pc}=1$, $\epsilon_{qc}=1$ \\ \hline $\epsilon_c=-1$, $\eta_c=-1$, $\epsilon_{pq}=1$, $\epsilon_{pc}=1$, $\epsilon_{qc}=1$ \end{tabular} &$ R_5 $&  $0$ & $2\mathbb{Z}$ & $0$ & $\mathbb{Z}_2$ & $\mathbb{Z}_2$ & $\mathbb{Z}$ & $0$ & $0$\\
      \hline
    PQC6& \begin{tabular}{c} $\epsilon_c=1$, $\eta_c=-1$, $\epsilon_{pq}=1$, $\epsilon_{pc}=1$, $\epsilon_{qc}=-1$ \\ \hline $\epsilon_c=-1$, $\eta_c=-1$, $\epsilon_{pq}=1$, $\epsilon_{pc}=1$, $\epsilon_{qc}=-1$ \end{tabular} & $ R_1 $&   $\mathbb{Z}_2$ & $\mathbb{Z}$ & $0$ & $0$ & $0$ & $2\mathbb{Z}$ & $0$ & $\mathbb{Z}_2$\\
      \hline
    PQC7& \begin{tabular}{c} $\epsilon_c=1$, $\eta_c=-1$, $\epsilon_{pq}=-1$, $\epsilon_{pc}=-1$, $\epsilon_{qc}=1$ \\ \hline $\epsilon_c=-1$, $\eta_c=1$, $\epsilon_{pq}=-1$, $\epsilon_{pc}=-1$, $\epsilon_{qc}=-1$ \end{tabular} & $ R_4 ^2 $&  $2\mathbb{Z}\oplus 2\mathbb{Z}$ & $0\oplus 0$ & $\mathbb{Z}_2 \oplus \mathbb{Z}_2$ & $\mathbb{Z}_2 \oplus \mathbb{Z}_2$ & $\mathbb{Z} \oplus \mathbb{Z}$ & $0\oplus 0$ & $0\oplus 0$ & $0\oplus 0$\\
      \hline
    PQC8& \begin{tabular}{c} $\epsilon_c=1$, $\eta=-1$, $\epsilon_{pq}=-1$, $\epsilon_{pc}=-1$, $\epsilon_{qc}=-1$ \\ \hline $\epsilon_c=-1$, $\eta_c=1$, $\epsilon_{pq}=-1$, $\epsilon_{pc}=-1$, $\epsilon_{qc}=1$ \end{tabular} & $ R_2 ^2 $&  $\mathbb{Z}_2 \oplus \mathbb{Z}_2$ & $\mathbb{Z}_2 \oplus \mathbb{Z}_2$ & $\mathbb{Z} \oplus \mathbb{Z}$ & $0\oplus 0$ & $0\oplus 0$ & $0\oplus 0$ & $2\mathbb{Z}\oplus 2\mathbb{Z}$ & $0\oplus 0$\\
      \hline
    PQC9& \begin{tabular}{c} $\epsilon_c=1$, $\eta_c=1$, $\epsilon_{pq}=1$, $\epsilon_{pc}=-1$, $\epsilon_{qc}=1$ \\ \hline $\epsilon_c=-1$, $\eta_c=-1$, $\epsilon_{pq}=1$, $\epsilon_{pc}=-1$, $\epsilon_{qc}=1$ \\ \hline $\epsilon_c=1$, $\eta_c=1$, $\epsilon_{pq}=1$, $\epsilon_{pc}=-1$, $\epsilon_{qc}=-1$ \\ \hline $\epsilon_c=-1$, $\eta_c=-1$, $\epsilon_{pq}=1$, $\epsilon_{pc}=-1$, $\epsilon_{qc}=-1$ \end{tabular} & $ R_7 $&  $0$ & $0$ & $0$ & $2\mathbb{Z}$ & $0$ & $\mathbb{Z}_2$ & $\mathbb{Z}_2$ & $\mathbb{Z}$\\
    \hline
    PQC10& \begin{tabular}{c} $\epsilon_c=1$, $\eta_c=1$, $\epsilon_{pq}=-1$, $\epsilon_{pc}=1$, $\epsilon_{qc}=1$ \\ \hline $\epsilon_c=-1$, $\eta_c=1$, $\epsilon_{pq}=-1$, $\epsilon_{pc}=1$, $\epsilon_{qc}=-1$ \\ \hline $\epsilon_c=1$, $\eta_c=1$, $\epsilon_{pq}=-1$, $\epsilon_{pc}=1$, $\epsilon_{qc}=-1$ \\ \hline $\epsilon_c=-1$, $\eta_c=1$, $\epsilon_{pq}=-1$, $\epsilon_{pc}=1$, $\epsilon_{qc}=1$ \end{tabular} & $ C_0 $& $\mathbb{Z}$ & $0$ & $\mathbb{Z}$ & $0$ & $\mathbb{Z}$ & $0$ & $\mathbb{Z}$ & $0$\\
    \hline
    PQC11& \begin{tabular}{c}  $\epsilon_c=1$, $\eta_c=-1$, $\epsilon_{pq}=1$, $\epsilon_{pc}=-1$, $\epsilon_{qc}=1$ \\ \hline $\epsilon_c=-1$, $\eta_c=1$, $\epsilon_{pq}=1$, $\epsilon_{pc}=-1$, $\epsilon_{qc}=1$  \\ \hline $\epsilon_c=1$, $\eta_c=-1$, $\epsilon_{pq}=1$, $\epsilon_{pc}=-1$, $\epsilon_{qc}=-1$ \\ \hline $\epsilon_c=-1$, $\eta_c=1$, $\epsilon_{pq}=1$, $\epsilon_{pc}=-1$, $\epsilon_{qc}=-1$ \end{tabular} & $ R_3 $& $0$ & $\mathbb{Z}_2$ & $\mathbb{Z}_2$ & $\mathbb{Z}$ & $0$ & $0$ & $0$& $2\mathbb{Z}$ \\
    \hline
    PQC12& \begin{tabular}{c} $\epsilon_c=1$, $\eta_c=-1$, $\epsilon_{pq}=-1$, $\epsilon_{pc}=1$, $\epsilon_{qc}=1$ \\ \hline $\epsilon_c=-1$, $\eta_c=-1$, $\epsilon_{pq}=-1$, $\epsilon_{pc}=1$, $\epsilon_{qc}=-1$ \\ \hline $\epsilon_c=1$, $\eta_c=-1$, $\epsilon_{pq}=-1$, $\epsilon_{pc}=1$, $\epsilon_{qc}=-1$ \\ \hline $\epsilon_c=-1$, $\eta_c=-1$, $\epsilon_{pq}=-1$, $\epsilon_{pc}=1$, $\epsilon_{qc}=1$ \end{tabular} & $ C_0 $& $\mathbb{Z}$ & $0$ & $\mathbb{Z}$ & $0$ & $\mathbb{Z}$ & $0$ & $\mathbb{Z}$ & $0$\\
    \hline

  \end{tabular}
  \end{center}
\end{table*}

\begin{table*}[htbp]\footnotesize
  \begin{center}
  \caption{\label{tab:tableV}  Periodic table of real gap classification of topological defects in non-Hermitian systems.  The rows correspond to the different GBL symmetry
  classes, while the columns depend
  on $\delta = d-D$.  For classes with P and at least one of Q and K symmetries, we omitted the signs of $\epsilon_q=1$ or $\epsilon_k=1$ or $\epsilon_q=\epsilon_k=1$. Also we omitted the classes with P and at least one of Q and K symmetries and $\epsilon_q=-1$ or $\epsilon_k=-1$, because they are equivalent to the corresponding classes with P and at least one of Q and K symmetries and $\epsilon_q=1$ or $\epsilon_k=1$.
  The topological numbers in the table are stable strong topological numbers.}

\begin{tabular}{|c|c|c|cccccccc|}
    \hline
    GBL&Gen. Rel.&Cl&$\delta=0$&1&2\;\;\;&\;\;\;3\;\;\;&\;\;\;4\;\;\;&\;\;\;\;5\;\;\;\;&\;\;\;\;6\;\;\;\;&\;\;\;\;7\;\;\;\\
  \hline
Non&  &$C_0$&$\mathbb{Z}$ & $0$ & $\mathbb{Z}$ & $0$ & $\mathbb{Z}$ & $0$ & $\mathbb{Z}$ & $0$\\
  \hline
P&   &$C_1  $&$0$ & $\mathbb{Z}$ & $0$ & $\mathbb{Z}$ & $0$ & $\mathbb{Z}$ & $0$ & $\mathbb{Z}$\\
  \hline
Qa& $\epsilon_q=1 $ & $ C_0^2 $ & $\mathbb{Z} \oplus \mathbb{Z}$ & $0\oplus 0$ & $\mathbb{Z} \oplus \mathbb{Z}$ & $0\oplus 0$ & $\mathbb{Z} \oplus \mathbb{Z}$ & $0\oplus 0$ & $\mathbb{Z} \oplus \mathbb{Z}$ &  $0\oplus 0$ \\
  \hline
  Qb& $\epsilon_q=-1 $ & $ C_1 $& $0$ & $\mathbb{Z}$ & $0$ & $\mathbb{Z}$ & $0$ & $\mathbb{Z}$ & $0$ & $\mathbb{Z}$\\
  \hline
K1a& $\epsilon_k =1$, $\eta_k =1$ & $ R_0 $ &  $\mathbb{Z}$ & $0$ & $0$ & $0$ & $2\mathbb{Z}$ & $0$ & $\mathbb{Z}_2$ & $\mathbb{Z}_2$\\
  \hline
  K1b& $\epsilon_k =-1$, $\eta_k =1$ & $ R_2 $ &  $\mathbb{Z}_2$ & $\mathbb{Z}_2$ & $\mathbb{Z}$ & $0$ & $0$ & $0$ & $2\mathbb{Z}$ & $0$\\
  \hline

K2a& $\epsilon_k =1$, $\eta_k =-1$ & $ R_4 $& $2\mathbb{Z}$ & $0$ & $\mathbb{Z}_2$ & $\mathbb{Z}_2$ & $\mathbb{Z}$ & $0$ & $0$ & $0$\\
  \hline
  K2b& $\epsilon_k =-1$,  $\eta_k =-1$ &$ R_6 $& $0$ & $0$ & $2\mathbb{Z}$ & $0$ & $\mathbb{Z}_2$ & $\mathbb{Z}_2$ & $\mathbb{Z}$ & $0$\\
  \hline

C1& $\epsilon_c=1$, $\eta_c=1$ & $ R_0 $& $\mathbb{Z}$ & $0$ & $0$ & $0$ & $2\mathbb{Z}$ & $0$ & $\mathbb{Z}_2$ & $\mathbb{Z}_2$\\
  \hline
C2& $\epsilon_c=1$, $\eta_c=-1$ & $ R_4 $& $2\mathbb{Z}$ & $0$ & $\mathbb{Z}_2$ & $\mathbb{Z}_2$ & $\mathbb{Z}$ & $0$ & $0$ & $0$\\
  \hline
C3& $\epsilon_c=-1$, $\eta_c=1$  & $ R_2 $& $\mathbb{Z}_2$ & $\mathbb{Z}_2$ & $\mathbb{Z}$ & $0$ & $0$ & $0$ & $2\mathbb{Z}$ & $0$\\
  \hline
C4& $\epsilon_c=-1$, $\eta_c=-1$ & $ R_6 $& $0$ & $0$ & $2\mathbb{Z}$ & $0$ & $\mathbb{Z}_2$ & $\mathbb{Z}_2$ & $\mathbb{Z}$ & $0$\\
  \hline

PQ1& $\epsilon_{pq}=1$ & $ C_1^2 $& $0\oplus 0$ & $\mathbb{Z} \oplus \mathbb{Z}$ & $0\oplus 0$ & $\mathbb{Z} \oplus \mathbb{Z}$ & $0\oplus 0$ & $\mathbb{Z} \oplus \mathbb{Z}$ & $0\oplus 0$ & $\mathbb{Z} \oplus \mathbb{Z}$\\
  \hline
PQ2& $\epsilon_{pq}=-1$ & $ C_0  $&  $\mathbb{Z}$ & $0$ & $\mathbb{Z}$ & $0$ & $\mathbb{Z}$ & $0$ & $\mathbb{Z}$ & $0$\\
  \hline
PK1& $\eta_k=1$, $\epsilon_{pk}=1$ & $ R_1  $&  $\mathbb{Z}_2$ & $\mathbb{Z}$ & $0$ & $0$ & $0$ & $2\mathbb{Z}$ & $0$ & $\mathbb{Z}_2$\\
  \hline
PK2& $\eta_k=-1$, $\epsilon_{pk}=1$ & $ R_5  $& $0$ & $2\mathbb{Z}$ & $0$ & $\mathbb{Z}_2$ & $\mathbb{Z}_2$ & $\mathbb{Z}$ & $0$ & $0$\\
  \hline
PK3a& \begin{tabular}{c} $\eta_k=1$, $\epsilon_{pk}=-1$  \end{tabular} & $ R_7 $& $0$ & $0$ & $0$ & $2\mathbb{Z}$ & $0$ & $\mathbb{Z}_2$ & $\mathbb{Z}_2$ & $\mathbb{Z}$\\
  \hline
PK3b& \begin{tabular}{c} $\eta_k=-1$, $\epsilon_{pk}=-1$   \end{tabular} & $ R_3 $&  $0$ & $\mathbb{Z}_2$ & $\mathbb{Z}_2$ & $\mathbb{Z}$ & $0$ & $0$ & $0$& $2\mathbb{Z}$ \\
  \hline
PC1& \begin{tabular}{c} $\epsilon_c=1$, $\eta_c=1$, $\epsilon_{pc}=1$ \\ \hline $\epsilon_c=-1$, $\eta_c=1$, $\epsilon_{pc}=1$ \end{tabular} & $ R_1 $& $\mathbb{Z}_2$ & $\mathbb{Z}$ & $0$ & $0$ & $0$ & $2\mathbb{Z}$ & $0$ & $\mathbb{Z}_2$\\
  \hline
PC2& \begin{tabular}{c} $\epsilon_c=1$, $\eta_c=1$, $\epsilon_{pc}=-1$ \\ \hline $\epsilon_c=-1$, $\eta_c=-1$, $\epsilon_{pc}=-1$ \end{tabular} & $ R_7 $& $0$ & $0$ & $0$ & $2\mathbb{Z}$ & $0$ & $\mathbb{Z}_2$ & $\mathbb{Z}_2$ & $\mathbb{Z}$\\
  \hline
PC3& \begin{tabular}{c} $\epsilon_c=1$, $\eta_c=-1$, $\epsilon_{pc}=1$ \\ \hline $\epsilon_c=-1$, $\eta_c=-1$, $\epsilon_{pc}=1$ \end{tabular} & $ R_5 $& $0$ & $2\mathbb{Z}$ & $0$ & $\mathbb{Z}_2$ & $\mathbb{Z}_2$ & $\mathbb{Z}$ & $0$ & $0$\\
  \hline
PC4& \begin{tabular}{c} $\epsilon_c=1$, $\eta_c=-1$, $\epsilon_{pc}=-1$ \\ \hline $\epsilon_c=-1$, $\eta_c=1$, $\epsilon_{pc}=-1$ \end{tabular} & $ R_3 $&  $0$ & $\mathbb{Z}_2$ & $\mathbb{Z}_2$ & $\mathbb{Z}$ & $0$ & $0$ & $0$& $2\mathbb{Z}$ \\
  \hline
QC1a& $\epsilon_q=1$, $\epsilon_c=1$, $\eta_c=1$, $\epsilon_{qc}=1$ & $ R_0^2 $&  $\mathbb{Z}\oplus \mathbb{Z}$ & $0\oplus 0$ & $0\oplus 0$ & $0\oplus 0$ & $2\mathbb{Z}\oplus 2\mathbb{Z}$ & $0\oplus 0$ & $\mathbb{Z}_2 \oplus \mathbb{Z}_2$ & $\mathbb{Z}_2 \oplus \mathbb{Z}_2$ \\
  \hline
  QC1b& $\epsilon_q=-1$, $\epsilon_c=1$, $\eta_c=1$, $\epsilon_{qc}=1$ & $ R_1 $&  $\mathbb{Z}_2$ & $\mathbb{Z}$ & $0$ & $0$ & $0$ & $2\mathbb{Z}$ & $0$ & $\mathbb{Z}_2$\\
  \hline

QC2a&  $\epsilon_q=1$, $\epsilon_c=1$, $\eta_c=1$, $\epsilon_{qc}=-1$  & $ C_0 $& $\mathbb{Z}$ & $0$ & $\mathbb{Z}$ & $0$ & $\mathbb{Z}$ & $0$ & $\mathbb{Z}$ & $0$\\
  \hline
  QC2b&  $\epsilon_q=-1$, $\epsilon_c=1$, $\eta_c=1$, $\epsilon_{qc}=-1$ & $ R_7 $&  $0$ & $0$ & $0$ & $2\mathbb{Z}$ & $0$ & $\mathbb{Z}_2$ & $\mathbb{Z}_2$ & $\mathbb{Z}$\\
  \hline

QC3a&  $\epsilon_q=1$, $\epsilon_c=1$, $\eta_c=-1$, $\epsilon_{qc}=1$  & $ R_4^2 $& $2\mathbb{Z}\oplus 2\mathbb{Z}$ & $0\oplus 0$ & $\mathbb{Z}_2 \oplus \mathbb{Z}_2$ & $\mathbb{Z}_2 \oplus \mathbb{Z}_2$ & $\mathbb{Z} \oplus \mathbb{Z}$ & $0\oplus 0$ & $0\oplus 0$ & $0\oplus 0$\\
  \hline
  QC3b&  $\epsilon_q=-1$, $\epsilon_c=1$, $\eta_c=-1$, $\epsilon_{qc}=1$  & $ R_5 $& $0$ & $2\mathbb{Z}$ & $0$ & $\mathbb{Z}_2$ & $\mathbb{Z}_2$ & $\mathbb{Z}$ & $0$ & $0$\\
  \hline

QC4a& $\epsilon_q=1$, $\epsilon_c=1$, $\eta_c=-1$, $\epsilon_{qc}=-1$  & $ C_0 $&  $\mathbb{Z}$ & $0$ & $\mathbb{Z}$ & $0$ & $\mathbb{Z}$ & $0$ & $\mathbb{Z}$ & $0$\\
  \hline
  QC4b& $\epsilon_q=-1$, $\epsilon_c=1$, $\eta_c=-1$, $\epsilon_{qc}=-1$  & $ R_3 $&  $0$ & $\mathbb{Z}_2$ & $\mathbb{Z}_2$ & $\mathbb{Z}$ & $0$ & $0$ & $0$& $2\mathbb{Z}$ \\
  \hline

QC5a& $\epsilon_q=1$, $\epsilon_c=-1$, $\eta_c=1$, $\epsilon_{qc}=1$  & $ R_2^2 $&  $\mathbb{Z}_2 \oplus \mathbb{Z}_2$ & $\mathbb{Z}_2 \oplus \mathbb{Z}_2$ & $\mathbb{Z} \oplus \mathbb{Z}$ & $0\oplus 0$ & $0\oplus 0$ & $0\oplus 0$ & $2\mathbb{Z}\oplus 2\mathbb{Z}$ & $0\oplus 0$\\
  \hline
  QC5b& $\epsilon_q=-1$, $\epsilon_c=-1$, $\eta_c=1$, $\epsilon_{qc}=1$  & $ R_1 $&   $\mathbb{Z}_2$ & $\mathbb{Z}$ & $0$ & $0$ & $0$ & $2\mathbb{Z}$ & $0$ & $\mathbb{Z}_2$\\
  \hline

QC6a& $\epsilon_q=1$, $\epsilon_c=-1$, $\eta_c=1$, $\epsilon_{qc}=-1$  & $ C_0 $& $\mathbb{Z}$ & $0$ & $\mathbb{Z}$ & $0$ & $\mathbb{Z}$ & $0$ & $\mathbb{Z}$ & $0$\\
  \hline
  QC6b& $\epsilon_q=-1$, $\epsilon_c=-1$, $\eta_c=1$, $\epsilon_{qc}=-1$  & $ R_3 $&  $0$ & $\mathbb{Z}_2$ & $\mathbb{Z}_2$ & $\mathbb{Z}$ & $0$ & $0$ & $0$& $2\mathbb{Z}$ \\
  \hline

QC7a& $\epsilon_q=1$, $\epsilon_c=-1$, $\eta_c=-1$, $\epsilon_{qc}=1$  & $ R_6^2 $& $0\oplus 0$ & $0\oplus 0$ & $2\mathbb{Z}\oplus 2\mathbb{Z}$ & $0\oplus 0$ & $\mathbb{Z}_2 \oplus \mathbb{Z}_2$ &$\mathbb{Z}_2 \oplus \mathbb{Z}_2$ & $\mathbb{Z} \oplus \mathbb{Z}$ & $0\oplus 0$\\
  \hline
  QC7b& $\epsilon_q=-1$, $\epsilon_c=-1$, $\eta_c=-1$, $\epsilon_{qc}=1$  & $ R_5 $& $0$ & $2\mathbb{Z}$ & $0$ & $\mathbb{Z}_2$ & $\mathbb{Z}_2$ & $\mathbb{Z}$ & $0$ & $0$\\
  \hline

QC8a& $\epsilon_q=1$, $\epsilon_c=-1$, $\eta_c=-1$, $\epsilon_{qc}=-1$   & $ C_0 $&  $\mathbb{Z}$ & $0$ & $\mathbb{Z}$ & $0$ & $\mathbb{Z}$ & $0$ & $\mathbb{Z}$ & $0$\\
  \hline
  QC8b& $\epsilon_q=-1$, $\epsilon_c=-1$, $\eta_c=-1$, $\epsilon_{qc}=-1$   & $ R_7 $&  $0$ & $0$ & $0$ & $2\mathbb{Z}$ & $0$ & $\mathbb{Z}_2$ & $\mathbb{Z}_2$ & $\mathbb{Z}$\\
  \hline

PQC1& \begin{tabular}{c} $\epsilon_c=1$, $\eta_c=1$, $\epsilon_{pq}=1$, $\epsilon_{pc}=1$, $\epsilon_{qc}=1$ \\ \hline $\epsilon_c=-1$, $\eta_c=1$, $\epsilon_{pq}=1$, $\epsilon_{pc}=1$, $\epsilon_{qc}=1$ \end{tabular} & $ R_1^2 $&  $\mathbb{Z}_2 \oplus \mathbb{Z}_2$ & $\mathbb{Z} \oplus \mathbb{Z}$ & $0 \oplus 0$ & $0 \oplus 0$ & $0 \oplus 0$ & $2\mathbb{Z}\oplus 2\mathbb{Z}$ & $0 \oplus 0$ & $\mathbb{Z}_2 \oplus \mathbb{Z}_2$\\
  \hline
PQC2& \begin{tabular}{c} $\epsilon_c=1$, $\eta_c=1$, $\epsilon_{pq}=1$, $\epsilon_{pc}=1$, $\epsilon_{qc}=-1$ \\ \hline $\epsilon_c=-1$, $\eta_c=1$, $\epsilon_{pq}=1$, $\epsilon_{pc}=1$, $\epsilon_{qc}=-1$ \end{tabular} & $ C_1 $&  $0$ & $\mathbb{Z}$ & $0$ & $\mathbb{Z}$ & $0$ & $\mathbb{Z}$ & $0$ & $\mathbb{Z}$\\
  \hline
PQC3& \begin{tabular}{c} $\epsilon_c=1$, $\eta_c=1$, $\epsilon_{pq}=-1$, $\epsilon_{pc}=-1$, $\epsilon_{qc}=1$ \\ \hline $\epsilon_c=-1$, $\eta_c=-1$, $\epsilon_{pq}=-1$, $\epsilon_{pc}=-1$, $\epsilon_{qc}=-1$ \end{tabular} & $ R_0 $&  $\mathbb{Z}$ & $0$ & $0$ & $0$ & $2\mathbb{Z}$ & $0$ & $\mathbb{Z}_2$ & $\mathbb{Z}_2$\\
  \hline
PQC4& \begin{tabular}{c} $\epsilon_c=1$, $\eta_c=1$, $\epsilon_{pq}=-1$, $\epsilon_{pc}=-1$, $\epsilon_{qc}=-1$ \\ \hline $\epsilon_c=-1$, $\eta_c=-1$, $\epsilon_{pq}=-1$, $\epsilon_{pc}=-1$, $\epsilon_{qc}=1$ \end{tabular} & $ R_6  $& $0$ & $0$ & $2\mathbb{Z}$ & $0$ & $\mathbb{Z}_2$ & $\mathbb{Z}_2$ & $\mathbb{Z}$ & $0$\\
  \hline
   \multicolumn{11}{c}{continued on next page}
  \end{tabular}
\end{center}
\end{table*}

\begin{table*}[htbp]\footnotesize
  \begin{center}
\begin{tabular}{|c|c|c|cccccccc|}
 \multicolumn{11}{c}{TABLE IV  --- continued} \\ \hline
PQC5& \begin{tabular}{c} $\epsilon_c=1$, $\eta_c=-1$, $\epsilon_{pq}=1$, $\epsilon_{pc}=1$, $\epsilon_{qc}=1$ \\ \hline $\epsilon_c=-1$, $\eta_c=-1$, $\epsilon_{pq}=1$, $\epsilon_{pc}=1$, $\epsilon_{qc}=1$ \end{tabular} &$ R_5^2 $&  $0\oplus 0$ & $2\mathbb{Z}\oplus 2\mathbb{Z}$ & $0\oplus 0$ & $\mathbb{Z}_2\oplus \mathbb{Z}_2$ & $\mathbb{Z}_2 \oplus \mathbb{Z}_2$ & $\mathbb{Z}\oplus \mathbb{Z}$ & $0\oplus 0$ & $0\oplus 0$\\
  \hline
PQC6& \begin{tabular}{c} $\epsilon_c=1$, $\eta_c=-1$, $\epsilon_{pq}=1$, $\epsilon_{pc}=1$, $\epsilon_{qc}=-1$ \\ \hline $\epsilon_c=-1$, $\eta_c=-1$, $\epsilon_{pq}=1$, $\epsilon_{pc}=1$, $\epsilon_{qc}=-1$ \end{tabular} & $ C_1 $&   $0$ & $\mathbb{Z}$ & $0$ & $\mathbb{Z}$ & $0$ & $\mathbb{Z}$ & $0$ & $\mathbb{Z}$\\
  \hline
PQC7& \begin{tabular}{c} $\epsilon_c=1$, $\eta_c=-1$, $\epsilon_{pq}=-1$, $\epsilon_{pc}=-1$, $\epsilon_{qc}=1$ \\ \hline $\epsilon_c=-1$, $\eta_c=1$, $\epsilon_{pq}=-1$, $\epsilon_{pc}=-1$, $\epsilon_{qc}=-1$ \end{tabular} & $ R_4  $&  $2\mathbb{Z}$ & $0$ & $\mathbb{Z}_2$ & $\mathbb{Z}_2$ & $\mathbb{Z}$ & $0$ & $0$ & $0$\\
  \hline
PQC8& \begin{tabular}{c} $\epsilon_c=1$, $\eta_c=-1$, $\epsilon_{pq}=-1$, $\epsilon_{pc}=-1$, $\epsilon_{qc}=-1$ \\ \hline $\epsilon_c=-1$, $\eta_c=1$, $\epsilon_{pq}=-1$, $\epsilon_{pc}=-1$, $\epsilon_{qc}=1$ \end{tabular} & $ R_2  $&  $\mathbb{Z}_2$ & $\mathbb{Z}_2$ & $\mathbb{Z}$ & $0$ & $0$ & $0$ & $2\mathbb{Z}$ & $0$\\
  \hline

PQC9a& \begin{tabular}{c} $\epsilon_c=1$, $\eta_c=1$, $\epsilon_{pq}=1$, $\epsilon_{pc}=-1$, $\epsilon_{qc}=1$ \\ \hline $\epsilon_c=-1$, $\eta_c=-1$, $\epsilon_{pq}=1$, $\epsilon_{pc}=-1$, $\epsilon_{qc}=1$  \end{tabular} & $ R_7^2 $&  $0\oplus 0$ & $0\oplus 0$ & $0\oplus 0$ & $2\mathbb{Z}\oplus 2\mathbb{Z}$ & $0\oplus 0$ & $\mathbb{Z}_2\oplus \mathbb{Z}_2$ &  $\mathbb{Z}_2\oplus \mathbb{Z}_2$ & $\mathbb{Z}\oplus \mathbb{Z}$\\
  \hline
  PQC9b& \begin{tabular}{c}  $\epsilon_c=1$, $\eta_c=1$, $\epsilon_{pq}=1$, $\epsilon_{pc}=-1$, $\epsilon_{qc}=-1$ \\ \hline $\epsilon_c=-1$, $\eta_c=-1$, $\epsilon_{pq}=1$, $\epsilon_{pc}=-1$, $\epsilon_{qc}=-1$ \end{tabular} & $ C_1 $&  $0$ & $\mathbb{Z}$ & $0$ & $\mathbb{Z}$ & $0$ & $\mathbb{Z}$ & $0$ & $\mathbb{Z}$\\
  \hline

PQC10a& \begin{tabular}{c} $\epsilon_c=1$, $\eta_c=1$, $\epsilon_{pq}=-1$, $\epsilon_{pc}=1$, $\epsilon_{qc}=1$ \\ \hline $\epsilon_c=-1$, $\eta_c=1$, $\epsilon_{pq}=-1$, $\epsilon_{pc}=1$, $\epsilon_{qc}=-1$  \end{tabular} & $ R_0 $& $\mathbb{Z}$ & $0$ & $0$ & $0$ & $2\mathbb{Z}$ & $0$ & $\mathbb{Z}_2$ & $\mathbb{Z}_2$\\
  \hline
  PQC10b& \begin{tabular}{c}  $\epsilon_c=1$, $\eta_c=1$, $\epsilon_{pq}=-1$, $\epsilon_{pc}=1$, $\epsilon_{qc}=-1$ \\ \hline $\epsilon_c=-1$, $\eta_c=1$, $\epsilon_{pq}=-1$, $\epsilon_{pc}=1$, $\epsilon_{qc}=1$ \end{tabular} & $ R_2 $&  $\mathbb{Z}_2$ & $\mathbb{Z}_2$ & $\mathbb{Z}$ & $0$ & $0$ & $0$ & $2\mathbb{Z}$ & $0$\\
  \hline

PQC11a& \begin{tabular}{c}$\epsilon_c=1$, $\eta_c=-1$, $\epsilon_{pq}=1$, $\epsilon_{pc}=-1$, $\epsilon_{qc}=1$ \\ \hline $\epsilon_c=-1$, $\eta_c=1$, $\epsilon_{pq}=1$, $\epsilon_{pc}=-1$, $\epsilon_{qc}=1$ \end{tabular} & $ R_3^2 $& $0\oplus 0$ & $\mathbb{Z}_2\oplus \mathbb{Z}_2$ & $\mathbb{Z}_2\oplus \mathbb{Z}_2$ & $\mathbb{Z}\oplus \mathbb{Z}$ & $0\oplus 0$ & $0\oplus 0$ & $0\oplus 0$& $2\mathbb{Z}\oplus 2\mathbb{Z}$ \\
  \hline
  PQC11b& \begin{tabular}{c}  $\epsilon_c=1$, $\eta_c=-1$, $\epsilon_{pq}=1$, $\epsilon_{pc}=-1$, $\epsilon_{qc}=-1$ \\ \hline $\epsilon_c=-1$, $\eta_c=1$, $\epsilon_{pq}=1$, $\epsilon_{pc}=-1$, $\epsilon_{qc}=-1$ \end{tabular} & $ C_1 $& $0$ & $\mathbb{Z}$ & $0$ & $\mathbb{Z}$ & $0$ & $\mathbb{Z}$ & $0$ & $\mathbb{Z}$\\
  \hline

PQC12a& \begin{tabular}{c} $\epsilon_c=1$, $\eta_c=-1$, $\epsilon_{pq}=-1$, $\epsilon_{pc}=1$, $\epsilon_{qc}=1$ \\ \hline $\epsilon_c=-1$, $\eta_c=-1$, $\epsilon_{pq}=-1$, $\epsilon_{pc}=1$, $\epsilon_{qc}=-1$  \end{tabular} & $ R_4 $& $2\mathbb{Z}$ & $0$ & $\mathbb{Z}_2$ & $\mathbb{Z}_2$ & $\mathbb{Z}$ & $0$ & $0$ & $0$\\
  \hline
  PQC12b& \begin{tabular}{c} $\epsilon_c=1$, $\eta_c=-1$, $\epsilon_{pq}=-1$, $\epsilon_{pc}=1$, $\epsilon_{qc}=-1$ \\ \hline $\epsilon_c=-1$, $\eta_c=-1$, $\epsilon_{pq}=-1$, $\epsilon_{pc}=1$, $\epsilon_{qc}=1$ \end{tabular} & $ R_6 $& $0$ & $0$ & $2\mathbb{Z}$ & $0$ & $\mathbb{Z}_2$ & $\mathbb{Z}_2$ & $\mathbb{Z}$ & $0$\\
  \hline
\end{tabular}

\end{center}
\end{table*}

\begin{table*}[h]\footnotesize
  \begin{center}
  \caption{\label{tab:table8}  Periodic table of imaginary gap classification of topological defects in non-Hermitian systems.  The rows correspond to different GBL symmetry
  classes, while the columns depend
  on $\delta = d-D$. For classes with P and at least one of Q and K symmetries, we omitted the signs of $\epsilon_q=1$ or $\epsilon_k=1$ or $\epsilon_q=\epsilon_k=1$. Also we omitted the classes with P and at least one of Q and K symmetries and $\epsilon_q=-1$ or $\epsilon_k=-1$, because they are equivalent to the corresponding classes with P and at least one of Q and K symmetries and $\epsilon_q=1$ or $\epsilon_k=1$. The topological numbers in the table are stable strong topological numbers. }

  \begin{tabular}{|c|c|c|cccccccc|}
    \hline
    GBL&Gen. Rel.&Cl&$\delta=0$&1&2\;\;\;&\;\;\;3\;\;\;&\;\;\;4\;\;\;&\;\;\;\;5\;\;\;\;&\;\;\;\;6\;\;\;\;&\;\;\;\;7\;\;\;\\
  \hline
Non&  &$C_0$&$\mathbb{Z}$ & $0$ & $\mathbb{Z}$ & $0$ & $\mathbb{Z}$ & $0$ & $\mathbb{Z}$ & $0$\\
  \hline
P&   &$C_1  $&$0$ & $\mathbb{Z}$ & $0$ & $\mathbb{Z}$ & $0$ & $\mathbb{Z}$ & $0$ & $\mathbb{Z}$\\
  \hline
Qa& $\epsilon_q=1$ & $ C_1 $& $0$ & $\mathbb{Z}$ & $0$ & $\mathbb{Z}$ & $0$ & $\mathbb{Z}$ & $0$ & $\mathbb{Z}$\\
  \hline
  Qb& $\epsilon_q=-1$ & $ C_0^2 $ & $\mathbb{Z} \oplus \mathbb{Z}$ & $0\oplus 0$ & $\mathbb{Z} \oplus \mathbb{Z}$ & $0\oplus 0$ & $\mathbb{Z} \oplus \mathbb{Z}$ & $0\oplus 0$ & $\mathbb{Z} \oplus \mathbb{Z}$ &  $0\oplus 0$ \\
  \hline

K1a& $\epsilon_k=1$ ,$\eta_k =1$ & $ R_2 $ &  $\mathbb{Z}_2$ & $\mathbb{Z}_2$ & $\mathbb{Z}$ & $0$ & $0$ & $0$ & $2\mathbb{Z}$ & $0$\\
  \hline
  K1b& $\epsilon_k=-1$ ,$\eta_k =1$ & $ R_0 $ &  $\mathbb{Z}$ & $0$ & $0$ & $0$ & $2\mathbb{Z}$ & $0$ & $\mathbb{Z}_2$ & $\mathbb{Z}_2$\\
  \hline

K2a& $\epsilon_k=1$ ,$\eta_k =-1$ & $ R_6 $& $0$ & $0$ & $2\mathbb{Z}$ & $0$ & $\mathbb{Z}_2$ & $\mathbb{Z}_2$ & $\mathbb{Z}$ & $0$\\
  \hline
  K2b& $\epsilon_k=-1$ , $\eta_k =-1$ & $ R_4 $& $2\mathbb{Z}$ & $0$ & $\mathbb{Z}_2$ & $\mathbb{Z}_2$ & $\mathbb{Z}$ & $0$ & $0$ & $0$\\
  \hline

C1& $\epsilon_c=1$, $\eta_c=1$ & $ R_0 $& $\mathbb{Z}$ & $0$ & $0$ & $0$ & $2\mathbb{Z}$ & $0$ & $\mathbb{Z}_2$ & $\mathbb{Z}_2$\\
  \hline
C2& $\epsilon_c=1$, $\eta_c=-1$ & $ R_4 $& $2\mathbb{Z}$ & $0$ & $\mathbb{Z}_2$ & $\mathbb{Z}_2$ & $\mathbb{Z}$ & $0$ & $0$ & $0$\\
  \hline
C3& $\epsilon_c=-1$, $\eta_c=1$ & $ R_2 $& $\mathbb{Z}_2$ & $\mathbb{Z}_2$ & $\mathbb{Z}$ & $0$ & $0$ & $0$ & $2\mathbb{Z}$ & $0$\\
  \hline
C4& $\epsilon_c=-1$, $\eta_c=-1$ & $ R_6 $& $0$ & $0$ & $2\mathbb{Z}$ & $0$ & $\mathbb{Z}_2$ & $\mathbb{Z}_2$ & $\mathbb{Z}$ & $0$\\
  \hline
PQ1& $\epsilon_{pq}=1$ & $ C_1^2 $& $0\oplus 0$ & $\mathbb{Z} \oplus \mathbb{Z}$ & $0\oplus 0$ & $\mathbb{Z} \oplus \mathbb{Z}$ & $0\oplus 0$ & $\mathbb{Z} \oplus \mathbb{Z}$ & $0\oplus 0$ & $\mathbb{Z} \oplus \mathbb{Z}$\\
  \hline
PQ2& $\epsilon_{pq}=-1$ & $ C_0  $&  $\mathbb{Z}$ & $0$ & $\mathbb{Z}$ & $0$ & $\mathbb{Z}$ & $0$ & $\mathbb{Z}$ & $0$\\
  \hline
PK1& $\eta_k=1$, $\epsilon_{pk}=1$ & $ R_1  $&  $\mathbb{Z}_2$ & $\mathbb{Z}$ & $0$ & $0$ & $0$ & $2\mathbb{Z}$ & $0$ & $\mathbb{Z}_2$\\
  \hline
PK2& $\eta_k=-1$, $\epsilon_{pk}=1$ & $ R_5  $& $0$ & $2\mathbb{Z}$ & $0$ & $\mathbb{Z}_2$ & $\mathbb{Z}_2$ & $\mathbb{Z}$ & $0$ & $0$\\
  \hline
PK3a& \begin{tabular}{c} $\eta_k=1$, $\epsilon_{pk}=-1$   \end{tabular} & $ R_3 $&  $0$ & $\mathbb{Z}_2$ & $\mathbb{Z}_2$ & $\mathbb{Z}$ & $0$ & $0$ & $0$& $2\mathbb{Z}$ \\
  \hline
  PK3b& \begin{tabular}{c} $\eta_k=-1$, $\epsilon_{pk}=-1$  \end{tabular} & $ R_7 $& $0$ & $0$ & $0$ & $2\mathbb{Z}$ & $0$ & $\mathbb{Z}_2$ & $\mathbb{Z}_2$ & $\mathbb{Z}$\\
  \hline
PC1& \begin{tabular}{c} $\epsilon_c=1$, $\eta_c=1$, $\epsilon_{pc}=1$ \\ \hline $\epsilon_c=-1$, $\eta_c=1$, $\epsilon_{pc}=1$ \end{tabular} & $ R_1 $& $\mathbb{Z}_2$ & $\mathbb{Z}$ & $0$ & $0$ & $0$ & $2\mathbb{Z}$ & $0$ & $\mathbb{Z}_2$\\
  \hline
PC2& \begin{tabular}{c} $\epsilon_c=1$, $\eta_c=1$, $\epsilon_{pc}=-1$ \\ \hline $\epsilon_c=-1$, $\eta_c=-1$, $\epsilon_{pc}=-1$ \end{tabular} & $ R_7 $& $0$ & $0$ & $0$ & $2\mathbb{Z}$ & $0$ & $\mathbb{Z}_2$ & $\mathbb{Z}_2$ & $\mathbb{Z}$\\
  \hline
PC3& \begin{tabular}{c} $\epsilon_c=1$, $\eta_c=-1$, $\epsilon_{pc}=1$ \\ \hline $\epsilon_c=-1$, $\eta_c=-1$, $\epsilon_{pc}=1$ \end{tabular} & $ R_5 $& $0$ & $2\mathbb{Z}$ & $0$ & $\mathbb{Z}_2$ & $\mathbb{Z}_2$ & $\mathbb{Z}$ & $0$ & $0$\\
  \hline
PC4& \begin{tabular}{c} $\epsilon_c=1$, $\eta_c=-1$, $\epsilon_{pc}=-1$ \\ \hline $\epsilon_c=-1$, $\eta_c=1$, $\epsilon_{pc}=-1$ \end{tabular} & $ R_3 $&  $0$ & $\mathbb{Z}_2$ & $\mathbb{Z}_2$ & $\mathbb{Z}$ & $0$ & $0$ & $0$& $2\mathbb{Z}$ \\
  \hline

QC1a& $\epsilon_q=1$ , $\epsilon_c=1$, $\eta_c=1$, $\epsilon_{qc}=1$ & $ R_1 $&  $\mathbb{Z}_2$ & $\mathbb{Z}$ & $0$ & $0$ & $0$ & $2\mathbb{Z}$ & $0$ & $\mathbb{Z}_2$\\
  \hline
  QC1b& $\epsilon_q=-1$ , $\epsilon_c=1$, $\eta_c=1$, $\epsilon_{qc}=1$ & $ R_0^2 $&  $\mathbb{Z}\oplus \mathbb{Z}$ & $0\oplus 0$ & $0\oplus 0$ & $0\oplus 0$ & $2\mathbb{Z}\oplus 2\mathbb{Z}$ & $0\oplus 0$ & $\mathbb{Z}_2 \oplus \mathbb{Z}_2$ & $\mathbb{Z}_2 \oplus \mathbb{Z}_2$ \\
  \hline

 QC2a&  $\epsilon_q=1$, $\epsilon_c=1$, $\eta_c=1$, $\epsilon_{qc}=-1$ & $ R_7 $&  $0$ & $0$ & $0$ & $2\mathbb{Z}$ & $0$ & $\mathbb{Z}_2$ & $\mathbb{Z}_2$ & $\mathbb{Z}$\\
  \hline
QC2b&  $\epsilon_q=-1$, $\epsilon_c=1$, $\eta_c=1$, $\epsilon_{qc}=-1$  & $ C_0 $& $\mathbb{Z}$ & $0$ & $\mathbb{Z}$ & $0$ & $\mathbb{Z}$ & $0$ & $\mathbb{Z}$ & $0$\\
  \hline

  QC3a&  $\epsilon_q=1$, $\epsilon_c=1$, $\eta_c=-1$, $\epsilon_{qc}=1$  & $ R_5 $& $0$ & $2\mathbb{Z}$ & $0$ & $\mathbb{Z}_2$ & $\mathbb{Z}_2$ & $\mathbb{Z}$ & $0$ & $0$\\
  \hline
QC3b&  $\epsilon_q=-1$, $\epsilon_c=1$, $\eta_c=-1$, $\epsilon_{qc}=1$  & $ R_4^2 $& $2\mathbb{Z}\oplus 2\mathbb{Z}$ & $0\oplus 0$ & $\mathbb{Z}_2 \oplus \mathbb{Z}_2$ & $\mathbb{Z}_2 \oplus \mathbb{Z}_2$ & $\mathbb{Z} \oplus \mathbb{Z}$ & $0\oplus 0$ & $0\oplus 0$ & $0\oplus 0$\\
  \hline

  QC4a& $\epsilon_q=1$, $\epsilon_c=1$, $\eta_c=-1$, $\epsilon_{qc}=-1$  & $ R_3 $&  $0$ & $\mathbb{Z}_2$ & $\mathbb{Z}_2$ & $\mathbb{Z}$ & $0$ & $0$ & $0$& $2\mathbb{Z}$ \\
  \hline
QC4b& $\epsilon_q=-1$, $\epsilon_c=1$, $\eta_c=-1$, $\epsilon_{qc}=-1$  & $ C_0 $&  $\mathbb{Z}$ & $0$ & $\mathbb{Z}$ & $0$ & $\mathbb{Z}$ & $0$ & $\mathbb{Z}$ & $0$\\
  \hline

  QC5a& $\epsilon_q=1$, $\epsilon_c=-1$, $\eta_c=1$, $\epsilon_{qc}=1$  & $ R_1 $&   $\mathbb{Z}_2$ & $\mathbb{Z}$ & $0$ & $0$ & $0$ & $2\mathbb{Z}$ & $0$ & $\mathbb{Z}_2$\\
  \hline
QC5b& $\epsilon_q=-1$, $\epsilon_c=-1$, $\eta_c=1$, $\epsilon_{qc}=1$  & $ R_2^2 $&  $\mathbb{Z}_2 \oplus \mathbb{Z}_2$ & $\mathbb{Z}_2 \oplus \mathbb{Z}_2$ & $\mathbb{Z} \oplus \mathbb{Z}$ & $0\oplus 0$ & $0\oplus 0$ & $0\oplus 0$ & $2\mathbb{Z}\oplus 2\mathbb{Z}$ & $0\oplus 0$\\
  \hline

  QC6a& $\epsilon_q=1$, $\epsilon_c=-1$, $\eta_c=1$, $\epsilon_{qc}=-1$  & $ R_3 $&  $0$ & $\mathbb{Z}_2$ & $\mathbb{Z}_2$ & $\mathbb{Z}$ & $0$ & $0$ & $0$& $2\mathbb{Z}$ \\
  \hline
QC6b& $\epsilon_q=-1$, $\epsilon_c=-1$, $\eta_c=1$, $\epsilon_{qc}=-1$  & $ C_0 $& $\mathbb{Z}$ & $0$ & $\mathbb{Z}$ & $0$ & $\mathbb{Z}$ & $0$ & $\mathbb{Z}$ & $0$\\
  \hline

  QC7a& $\epsilon_q=1$, $\epsilon_c=-1$, $\eta_c=-1$, $\epsilon_{qc}=1$  & $ R_5 $& $0$ & $2\mathbb{Z}$ & $0$ & $\mathbb{Z}_2$ & $\mathbb{Z}_2$ & $\mathbb{Z}$ & $0$ & $0$\\
  \hline
QC7b& $\epsilon_q=-1$, $\epsilon_c=-1$, $\eta_c=-1$, $\epsilon_{qc}=1$  & $ R_6^2 $& $0\oplus 0$ & $0\oplus 0$ & $2\mathbb{Z}\oplus 2\mathbb{Z}$ & $0\oplus 0$ & $\mathbb{Z}_2 \oplus \mathbb{Z}_2$ &$\mathbb{Z}_2 \oplus \mathbb{Z}_2$ & $\mathbb{Z} \oplus \mathbb{Z}$ & $0\oplus 0$\\
  \hline

 QC8a& $\epsilon_q=1$, $\epsilon_c=-1$, $\eta_c=-1$, $\epsilon_{qc}=-1$   & $ R_7 $&  $0$ & $0$ & $0$ & $2\mathbb{Z}$ & $0$ & $\mathbb{Z}_2$ & $\mathbb{Z}_2$ & $\mathbb{Z}$\\
  \hline
QC8b& $\epsilon_q=-1$, $\epsilon_c=-1$, $\eta_c=-1$, $\epsilon_{qc}=-1$   & $ C_0 $&  $\mathbb{Z}$ & $0$ & $\mathbb{Z}$ & $0$ & $\mathbb{Z}$ & $0$ & $\mathbb{Z}$ & $0$\\
  \hline

PQC1& \begin{tabular}{c} $\epsilon_c=1$, $\eta_c=1$, $\epsilon_{pq}=1$, $\epsilon_{pc}=1$, $\epsilon_{qc}=1$ \\ \hline $\epsilon_c=-1$, $\eta_c=1$, $\epsilon_{pq}=1$, $\epsilon_{pc}=1$, $\epsilon_{qc}=1$ \end{tabular} & $ R_1^2 $&  $\mathbb{Z}_2 \oplus \mathbb{Z}_2$ & $\mathbb{Z} \oplus \mathbb{Z}$ & $0 \oplus 0$ & $0 \oplus 0$ & $0 \oplus 0$ & $2\mathbb{Z}\oplus 2\mathbb{Z}$ & $0 \oplus 0$ & $\mathbb{Z}_2 \oplus \mathbb{Z}_2$\\
  \hline
PQC2& \begin{tabular}{c} $\epsilon_c=1$, $\eta_c=1$, $\epsilon_{pq}=1$, $\epsilon_{pc}=1$, $\epsilon_{qc}=-1$ \\ \hline $\epsilon_c=-1$, $\eta_c=1$, $\epsilon_{pq}=1$, $\epsilon_{pc}=1$, $\epsilon_{qc}=-1$ \end{tabular} & $ C_1 $&  $0$ & $\mathbb{Z}$ & $0$ & $\mathbb{Z}$ & $0$ & $\mathbb{Z}$ & $0$ & $\mathbb{Z}$\\
  \hline
PQC3& \begin{tabular}{c} $\epsilon_c=1$, $\eta_c=1$, $\epsilon_{pq}=-1$, $\epsilon_{pc}=-1$, $\epsilon_{qc}=1$ \\ \hline $\epsilon_c=-1$, $\eta_c=-1$, $\epsilon_{pq}=-1$, $\epsilon_{pc}=-1$, $\epsilon_{qc}=-1$ \end{tabular} & $ R_0 $&  $\mathbb{Z}$ & $0$ & $0$ & $0$ & $2\mathbb{Z}$ & $0$ & $\mathbb{Z}_2$ & $\mathbb{Z}_2$\\
  \hline
PQC4& \begin{tabular}{c} $\epsilon_c=1$, $\eta_c=1$, $\epsilon_{pq}=-1$, $\epsilon_{pc}=-1$, $\epsilon_{qc}=-1$ \\ \hline $\epsilon_c=-1$, $\eta_c=-1$, $\epsilon_{pq}=-1$, $\epsilon_{pc}=-1$, $\epsilon_{qc}=1$ \end{tabular} & $ R_6  $& $0$ & $0$ & $2\mathbb{Z}$ & $0$ & $\mathbb{Z}_2$ & $\mathbb{Z}_2$ & $\mathbb{Z}$ & $0$\\
  \hline

  \multicolumn{11}{c}{continued on next page}
  \end{tabular}
\end{center}
\end{table*}

\begin{table*}[htbp]\footnotesize
  \begin{center}
\begin{tabular}{|c|c|c|cccccccc|}
 \multicolumn{11}{c}{TABLE V --- continued} \\ \hline

 PQC5& \begin{tabular}{c} $\epsilon_c=1$, $\eta_c=-1$, $\epsilon_{pq}=1$, $\epsilon_{pc}=1$, $\epsilon_{qc}=1$ \\ \hline $\epsilon_c=-1$, $\eta_c=-1$, $\epsilon_{pq}=1$, $\epsilon_{pc}=1$, $\epsilon_{qc}=1$ \end{tabular} &$ R_5^2 $&  $0\oplus 0$ & $2\mathbb{Z}\oplus 2\mathbb{Z}$ & $0\oplus 0$ & $\mathbb{Z}_2\oplus \mathbb{Z}_2$ & $\mathbb{Z}_2 \oplus \mathbb{Z}_2$ & $\mathbb{Z}\oplus \mathbb{Z}$ & $0\oplus 0$ & $0\oplus 0$\\
  \hline
PQC6& \begin{tabular}{c} $\epsilon_c=1$, $\eta_c=-1$, $\epsilon_{pq}=1$, $\epsilon_{pc}=1$, $\epsilon_{qc}=-1$ \\ \hline $\epsilon_c=-1$, $\eta_c=-1$, $\epsilon_{pq}=1$, $\epsilon_{pc}=1$, $\epsilon_{qc}=-1$ \end{tabular} & $ C_1 $&   $0$ & $\mathbb{Z}$ & $0$ & $\mathbb{Z}$ & $0$ & $\mathbb{Z}$ & $0$ & $\mathbb{Z}$\\
  \hline
PQC7& \begin{tabular}{c} $\epsilon_c=1$, $\eta_c=-1$, $\epsilon_{pq}=-1$, $\epsilon_{pc}=-1$, $\epsilon_{qc}=1$ \\ \hline $\epsilon_c=-1$, $\eta_c=1$, $\epsilon_{pq}=-1$, $\epsilon_{pc}=-1$, $\epsilon_{qc}=-1$ \end{tabular} & $ R_4  $&  $2\mathbb{Z}$ & $0$ & $\mathbb{Z}_2$ & $\mathbb{Z}_2$ & $\mathbb{Z}$ & $0$ & $0$ & $0$\\
  \hline
PQC8& \begin{tabular}{c} $\epsilon_c=1$, $\eta_c=-1$, $\epsilon_{pq}=-1$, $\epsilon_{pc}=-1$, $\epsilon_{qc}=-1$ \\ \hline $\epsilon_c=-1$, $\eta_c=1$, $\epsilon_{pq}=-1$, $\epsilon_{pc}=-1$, $\epsilon_{qc}=1$ \end{tabular} & $ R_2  $&  $\mathbb{Z}_2$ & $\mathbb{Z}_2$ & $\mathbb{Z}$ & $0$ & $0$ & $0$ & $2\mathbb{Z}$ & $0$\\
  \hline

PQC9a& \begin{tabular}{c} $\epsilon_c=1$, $\eta_c=1$, $\epsilon_{pq}=1$, $\epsilon_{pc}=-1$, $\epsilon_{qc}=1$ \\ \hline $\epsilon_c=-1$, $\eta_c=-1$, $\epsilon_{pq}=1$, $\epsilon_{pc}=-1$, $\epsilon_{qc}=1$  \end{tabular} & $ C_1 $&  $0$ & $\mathbb{Z}$ & $0$ & $\mathbb{Z}$ & $0$ & $\mathbb{Z}$ & $0$ & $\mathbb{Z}$\\
  \hline
  PQC9b& \begin{tabular}{c}  $\epsilon_c=1$, $\eta_c=1$, $\epsilon_{pq}=1$, $\epsilon_{pc}=-1$, $\epsilon_{qc}=-1$ \\ \hline $\epsilon_c=-1$, $\eta_c=-1$, $\epsilon_{pq}=1$, $\epsilon_{pc}=-1$, $\epsilon_{qc}=-1$ \end{tabular} & $ R_7^2 $&  $0\oplus 0$ & $0\oplus 0$ & $0\oplus 0$ & $2\mathbb{Z}\oplus 2\mathbb{Z}$ & $0\oplus 0$ & $\mathbb{Z}_2\oplus \mathbb{Z}_2$ &  $\mathbb{Z}_2\oplus \mathbb{Z}_2$ & $\mathbb{Z}\oplus \mathbb{Z}$\\
  \hline

PQC10a& \begin{tabular}{c} $\epsilon_c=1$, $\eta_c=1$, $\epsilon_{pq}=-1$, $\epsilon_{pc}=1$, $\epsilon_{qc}=1$ \\ \hline $\epsilon_c=-1$, $\eta_c=1$, $\epsilon_{pq}=-1$, $\epsilon_{pc}=1$, $\epsilon_{qc}=-1$  \end{tabular} &$ R_2 $&  $\mathbb{Z}_2$ & $\mathbb{Z}_2$ & $\mathbb{Z}$ & $0$ & $0$ & $0$ & $2\mathbb{Z}$ & $0$\\
  \hline
  PQC10b& \begin{tabular}{c}  $\epsilon_c=1$, $\eta_c=1$, $\epsilon_{pq}=-1$, $\epsilon_{pc}=1$, $\epsilon_{qc}=-1$ \\ \hline $\epsilon_c=-1$, $\eta_c=1$, $\epsilon_{pq}=-1$, $\epsilon_{pc}=1$, $\epsilon_{qc}=1$ \end{tabular} & $ R_0 $& $\mathbb{Z}$ & $0$ & $0$ & $0$ & $2\mathbb{Z}$ & $0$ & $\mathbb{Z}_2$ & $\mathbb{Z}_2$\\
  \hline

PQC11a& \begin{tabular}{c} $\epsilon_c=1$, $\eta_c=-1$, $\epsilon_{pq}=1$, $\epsilon_{pc}=-1$, $\epsilon_{qc}=1$ \\ \hline $\epsilon_c=-1$, $\eta_c=1$, $\epsilon_{pq}=1$, $\epsilon_{pc}=-1$, $\epsilon_{qc}=1$  \end{tabular} &$ C_1 $& $0$ & $\mathbb{Z}$ & $0$ & $\mathbb{Z}$ & $0$ & $\mathbb{Z}$ & $0$ & $\mathbb{Z}$\\
  \hline
  PQC11b& \begin{tabular}{c} $\epsilon_c=1$, $\eta_c=-1$, $\epsilon_{pq}=1$, $\epsilon_{pc}=-1$, $\epsilon_{qc}=-1$ \\ \hline $\epsilon_c=-1$, $\eta_c=1$, $\epsilon_{pq}=1$, $\epsilon_{pc}=-1$, $\epsilon_{qc}=-1$ \end{tabular} & $ R_3^2 $& $0\oplus 0$ & $\mathbb{Z}_2\oplus \mathbb{Z}_2$ & $\mathbb{Z}_2\oplus \mathbb{Z}_2$ & $\mathbb{Z}\oplus \mathbb{Z}$ & $0\oplus 0$ & $0\oplus 0$ & $0\oplus 0$& $2\mathbb{Z}\oplus 2\mathbb{Z}$ \\
  \hline

PQC12a& \begin{tabular}{c} $\epsilon_c=1$, $\eta_c=-1$, $\epsilon_{pq}=-1$, $\epsilon_{pc}=1$, $\epsilon_{qc}=1$ \\ \hline $\epsilon_c=-1$, $\eta_c=-1$, $\epsilon_{pq}=-1$, $\epsilon_{pc}=1$, $\epsilon_{qc}=-1$  \end{tabular} &$ R_6 $& $0$ & $0$ & $2\mathbb{Z}$ & $0$ & $\mathbb{Z}_2$ & $\mathbb{Z}_2$ & $\mathbb{Z}$ & $0$\\
\hline
  PQC12b& \begin{tabular}{c} $\epsilon_c=1$, $\eta_c=-1$, $\epsilon_{pq}=-1$, $\epsilon_{pc}=1$, $\epsilon_{qc}=-1$ \\ \hline $\epsilon_c=-1$, $\eta_c=-1$, $\epsilon_{pq}=-1$, $\epsilon_{pc}=1$, $\epsilon_{qc}=1$ \end{tabular} & $ R_4 $& $2\mathbb{Z}$ & $0$ & $\mathbb{Z}_2$ & $\mathbb{Z}_2$ & $\mathbb{Z}$ & $0$ & $0$ & $0$\\
  \hline
\end{tabular}
\end{center}
\end{table*}

    \clearpage

    \twocolumngrid

  \section{Construction of explicit topological invariants for point defects}

    Although the periodic tables of topological classification are given,  they do not provide explicit forms of topological invariants. In this section, we discuss the construction of explicit topological invariants for point gap systems by considering some specific examples. To begin with, we note that the open boundary condition in one dimension is the simplest example of topological defects. For this case, our topological classification of topological defects ($d=1, D=0$) is consistent with classification of one-dimensional gapped system ($d=1$). 
    Next we discuss several one-dimensional and two-dimensional examples.

    {\bf Class A (Non), Class AI (K1) and Class AII (K2).}  Given the Hamiltonian $H(\bf{k},\bf{r})$, which describes a point defect in $d$ dimensions and is a function of d momentum variables and $D=d-1$ position variables. For any invertible Hamiltonian, it has a polar decomposition $H=UP$, where $U$ is a unitary matrix and $P$ is a positive definite Hermitian matrix. Such a decomposition is unique, and the Hamiltonian $H(\bf{k},\bf{r})$ can be continuously deformed to $U(\bf{k},\bf{r})$ with the same symmetry constraint \cite{Zhou,Sato}. Since $H(\bf{k},\bf{r})$ is topologically equivalent to $U(\bf{k},\bf{r})$, the $\mathbb{Z}$ topological number is defined as:
    \begin{equation}
      n=\frac{(d-1)!}{(2d-1)!(2\pi i)^d} \int _{T^d\times S^{d-1}}Tr[(UdU^\dagger)^{2d-1}] ,
    \end{equation}
which is the winding number associated with the homotopy group $\pi_{2d-1}[U(n)]=\mathbb{Z}$ \cite{Teo}.

    {\bf Class P, Class PK1 and Class PK2.}  Suppose that the Hamiltonian is $H(\bf{k},\bf{r})$ and the system has $P$ symmetry described by $p=\sigma_z$.  The polar decomposition is $H=UP$. Consider the symmetry constraint $\sigma_z U= -U\sigma_z$ \cite{Zhou}, and $U$ can be represented as
    \begin{equation}
      U(\bf{k},\bf{r})=\left[ \begin{array}{cc}
          0 & U_1 (\bf{k},\bf{r})\\
          U_2 (\bf{k},\bf{r})& 0
          \end{array}
          \right ].  \label{PKA}
      \end{equation}
       Then we can define the $\mathbb{Z}\oplus \mathbb{Z}$ topological number as
        \begin{equation}
       n_j=\frac{(d-1)!}{(2d-1)!(2\pi i)^d} \int _{T^d\times S^{d-1}}Tr[(U_jdU_j^\dagger)^{2d-1}],
        \end{equation}
        where $j=1$ and $2$.

       {\bf Class PK3.}  Suppose that the Hamiltonian of a point defect is $H(\bf{k},\bf{r})$, and the system has both $P$ and $K$ symmetries described by $p=\sigma_z$ and $k=\sigma_x$, respectively. The polar decomposition is $H=UP$. Consider the symmetry constraint $\sigma_z U(\bf{k},\bf{r})$$= -U({\bf k,r})\sigma_z$ and $\sigma_x U^*({\bf -k,r})=U({\bf k,r}) \sigma_x$, and $U$ can be represented as
    \begin{equation}
      U(\bf{k},\bf{r})=\left[ \begin{array}{cc}
          0 & U_1 (\bf{k},\bf{r})\\
          U_1^* (\bf{-k},\bf{r})& 0
          \end{array}
          \right ] .  \label{PK3}
      \end{equation}
       Then we can define the $\mathbb{Z}$ topological number as
       \begin{equation} \label{Topnum}
       n=\frac{(d-1)!}{(2d-1)!(2\pi i)^d} \int _{T^d\times S^{d-1}}Tr[(U_1dU_1^\dagger)^{2d-1}].
       \end{equation}

       {\bf Class PQ1.} The Hamiltonian of a point defect is $H(\bf{k},\bf{r})$ with the $P$ and $Q$ symmetries described by $p=q=\sigma_z$. Consider the symmetry constraints $\sigma_z H(\bf{k},\bf{r})$$= -$$H(\bf{k},\bf{r})$$\sigma_z$ and $\sigma_z H^\dagger (\bf{k},\bf{r})$$= H(\bf{k},\bf{r})$$\sigma_z$ as well as the polar decomposition is $H=UP$, and thus $U$ can be written as
       \begin{equation}
        H(\bf{k},\bf{r})=\left[ \begin{array}{cc}
            0 & U_1 (\bf{k},\bf{r})\\
            -U_1^\dagger (\bf{k},\bf{r})& 0
            \end{array}
            \right ] .  \label{PQ1}
        \end{equation}
         Then we can define the $\mathbb{Z}$ topological number by Eq.({\ref{Topnum}}).

         {\bf Class PC1.} The Hamiltonian of a point defect is $H(\bf{k},\bf{r})$ with the $P$ symmetry described by $p=\sigma_z$ and $C$ symmetry described by  $c=\sigma_0$  as well as the polar decomposition $H=UP$. Consider the symmetry constraints $\sigma_z U(\bf{k},\bf{r})$$= -$$U(\bf{k},\bf{r})$$\sigma_z$ and $ U^T(\bf{-k},\bf{r})$$= U(\bf{k},\bf{r})$, and thus $U$ has the following form:
       \begin{equation}
        U(\bf{k},\bf{r})=\left[ \begin{array}{cc}
            0 & U_1({\bf k,r})\\
            U_1^T ({\bf -k,r})& 0
            \end{array}
            \right ] .  \label{PC1}
        \end{equation}
         Then we can define the $\mathbb{Z}$ topological number  by Eq.({\ref{Topnum}}).

         {\bf Class PC3.} The Hamiltonian of a point defect is $H(\bf{k},\bf{r})$ with $P$ symmetry described by $p=\sigma_z\otimes \sigma_0$ and $C$ symmetry described by $c=\sigma_0 \otimes \sigma_y$ as well as the polar decomposition $H=UP$. Consider the symmetry constraints $p U(\bf{k},\bf{r})$$= -$$U(\bf{k},\bf{r})$$p$ and $ c U^T(\bf{-k},\bf{r})$$= U(\bf{k},\bf{r})$$c$, and thus $U$ has the following form:
       \begin{equation}
        U(\bf{k},\bf{r})=\left[ \begin{array}{cc}
            0 & U_1({\bf k,r})\\
            \sigma_y U_1^T ({\bf-k,r}) \sigma _y& 0
            \end{array}
            \right ] .  \label{PC3}
        \end{equation}
         Then we can define the $\mathbb{Z}$ topological number  by Eq.({\ref{Topnum}}).

         {\bf Class PQC1 and Class QC5.} If we chose $q=\mathbb{I}$, the Hamiltonian reduces to a Hermitian Hamiltonian. And the non-Hermitian classes of PQC1 and QC5 reduce to Hermitian classes of BDI and D. So we can define the $\mathbb{Z}$ and $\mathbb{Z}_2$ topological numbers for classes of PQC1 and QC5 the same as the Hermitian classes of BDI and D \cite{Teo},  which is consistent with our classification.

           \section{examples of point-defect models}

           To gain an intuitive understanding of nontrivial topological defects in non-Hermitian systems, we construct some examples with a point gap. Through the study of some concrete models, we discuss the correspondence between topological number and zero modes at the defect.
           \subsection{Point defects in 1D PC1 class}
           Consider the one-dimensional (1D) lattice models described by:
           \begin{equation}
            H_1=\sum _n [t_1 (a_n ^\dagger b_n+b_n ^\dagger a_n) + t_2e^{i\theta _1}(b_n ^\dagger a_{n+1}+a_{n+1} ^\dagger b_n)]
            \end{equation}
            and
            \begin{equation}
            H_2=\sum _n [t_3 (a_n ^\dagger b_n+b_n ^\dagger a_n) + t_4e^{i\theta _2}(b_n ^\dagger a_{n+1}+a_{n+1} ^\dagger b_n)].
            \end{equation}
            These two Hamiltonians have the same forms but with different parameters. If we consider the periodic boundary condition, both of them can be represented in the momentum space after a Fourier transformation. The corresponding Hamiltonians in the $k$ space are
           \begin{equation}
            H_1(k)=\left[ \begin{array}{cc}
                0 & t_1+t_2e^{i\theta _1-ik}\\
                t_1+t_2e^{i\theta _1+ik} & 0
                \end{array}
                \right ]   \label{h1}
            \end{equation}
            and
            \begin{equation}
                H_2(k)=\left[ \begin{array}{cc}
                    0 & t_3+t_4e^{i\theta _2-ik}\\
                    t_3+t_4e^{i\theta _2+ik} & 0
                    \end{array}
                    \right ] .   \label{h2}
                \end{equation}
Both the Hamiltonians belong to the $PC1$ Class as they have $P$ and $C$ symmetries with $p=\sigma_z$ and $c=\sigma_0$. The topological properties of the systems are characterized by the $\mathbb{Z}$ topological numbers, i.e., $n_1= \frac{1}{2\pi i}\int _0 ^{2\pi} dk\partial_kln(t_1+t_2e^{i\theta _1-ik}$ and $n_2=\frac{1}{2\pi i}\int _0 ^{2\pi} dk\partial_kln(t_3+t_4e^{i\theta _2-ik})$ for $H_1$ and $H_2$, respectively.
                \begin{figure}[h] 
                  \includegraphics[width=3.3in]{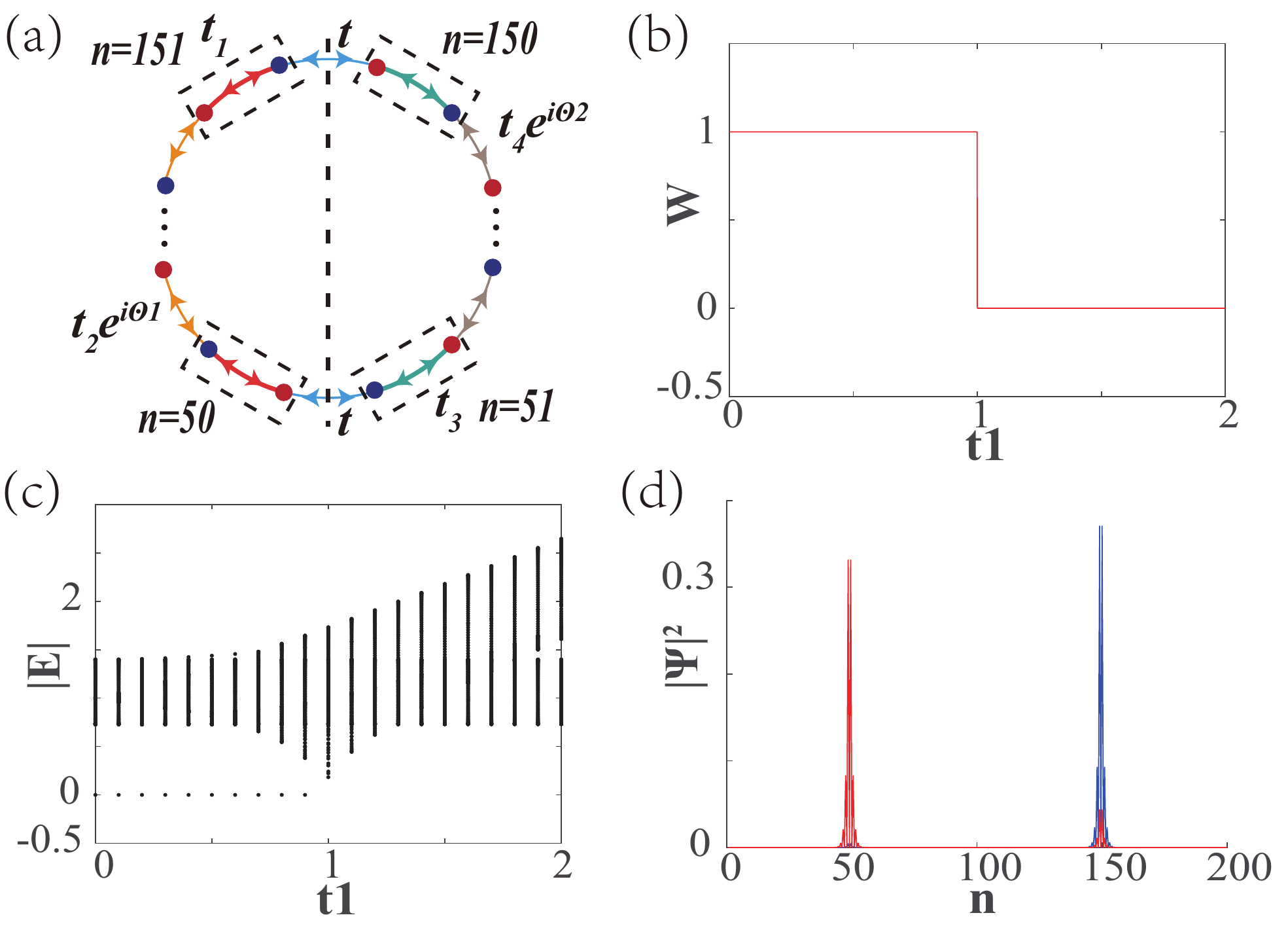}\\
                  \caption{ Parameters are set as: $t=t_2=t_3=1$, $t_4=0.5$, $\theta_1=\pi/3$ and $\theta_2=\pi/4$. The number of cells for $H_1$ is 100, and the number of cells for $H_2$ is also 100. (a) Schematic diagram, the left part represents $H_1$ and the right part represents $H_2$. (b) The topological number $W$ versus $t_1$. (c) Energy spectrum as a function of $t_1$. (d) The spatial distributions for zero modes of the system with $t_1=0.5$. There is a zero mode at the place of each connected point (defect).} \label{fig3}
                 \end{figure}

           Now we connect $H_1$ and $H_2$ together end to end to form a new Hamiltonian $H$. As schematically displayed in Fig.(\ref{fig3}a), $H_1$ and $H_2$ are coupled together with a coupling constant $t$. Each connected point can be viewed as a topological defect. The topological invariant for the defect is characterized by  $W=n_2-n_1$. In Fig.(\ref{fig3}b), we display $W$ versus $t_1$ by fixing $t=t_2=t_3=1$, $t_4=0.5$, $\theta_1=\pi/3$ and $\theta_2=\pi/4$. In the region of $t_1<1$, we have $W=1$, which indicates $H_1$ and $H_2$ in topologically different phases. Correspondingly, we observe the existence of zero mode states in the region of $t_1<1$ as shown in Fig.(\ref{fig3}c). The zero mode states are found to be localized at the places of defects, according to Fig.(\ref{fig3}d) for $t_1=0.5$.

           Next we consider the continuous model:
          \begin{equation}       \label{opc1}
           H(k,x)=\hat{k} \sigma_z +[ m(x)+i\alpha ] \sigma_x,
          \end{equation}
          where $\hat{k}$ represents the momentum operator ($\hat{k}=-i \partial_x$ in the coordinate representation), $\alpha$ is a real constant and $m(x)$ is a function of $x$, which changes sign as $x$ crosses the zero point.
          Here we take $|\alpha| \ll |m(x)|$ with $x \in S^0$, where $S^0$ is the $0D$ loop around the defect. When $\alpha=0$, this model reduces to the Jackiw-Rebbi model \cite{JRD1,Leykam}. The Hamiltonian has $P$ and $C$ symmetries with $\epsilon_c=-1$, $\eta_c=1$, $c=\sigma_z$ and $p=\sigma_y$, and belongs to the $PC1$ class. The topological defect is characterized by a $\mathbb{Z}$ topological number. When $\alpha$ continuously changes from $0$ to nonzero, $det(H)\ne 0$ on $S^0$ because $|\alpha| \ll |m(x)|$ on $S^0$. Then the topological number is the same as that of Jackiw-Rebbi model, i.e.,
           \begin{equation}
           W =\frac{1}{2}[ sgn(m(x_1))-sgn(m(x_2)) ],
           \end{equation}
           where $x_1>0$ and $x_2<0$ with $x_1,x_2 \in S^0$.

          When $m=b x$ ($b>0$ is a constant), 
          in the momentum representation, $\hat{x}= i\partial_k$ and the Hamiltonian can be written as:
          \begin{equation}
            H(k,i\partial_k)=k\sigma_z +(bi\partial_k+i\alpha)\sigma_x .
           \end{equation}
            Define $H'=e^{\alpha k/b}He^{-\alpha k/b}$, then $H'$ is given by
            \begin{equation}
              H'(\hat{k},\hat{x})=\hat{k}\sigma_z +b\hat{x}\sigma_x,
            \end{equation}
          which is a special form of  Jackiw-Rebbi Hamiltonian and has a topologically protected zero mode at the defect. Denote the zero mode state of $H'$  as $\Phi '_0(x)=\langle x| \Phi '_0 \rangle$, then $H$ has a topologically protected zero mode $\Phi_0(x)=e^{i\alpha \partial_x/b}\Phi '_0(x)$ at the defect. Our results show that there is a correspondence between zero modes of defect and the topological number of defect.

           For the non-Hermitian continuous model described by
          \begin{equation}       \label{opc1}
           H(k,x)= (-i\partial_x +i \beta) \sigma_z + m(x) \sigma_x,
          \end{equation}
         we can make a similarity transformation  $H'=e^{-\beta x}H e^{\beta x}$ and get
          \begin{equation}
              H'(\hat{k},\hat{x})=-i\partial_x \sigma_z + m(x) \sigma_x,
            \end{equation}
   which is a standard Jackiw-Rebbi Hamiltonian. Similarly,  we conclude that $H$ has a topologically protected zero mode $\Phi_0(x)=e^{\beta x} \Phi '_0(x)$, where $\Phi '_0(x)$ is the zero mode wavefunction of $H'$.
   
    \subsection{Point defects in 1D K1 class}

              Consider a toy model described by
              \begin{equation}
              H=\left\{
\begin{aligned}
e^{i\hat{k}},  ~~~~~x<0 \\
1        ,   ~~~~~x>0
\end{aligned}
\right.
\end{equation}
 where $\hat{k}$ represents the momentum operator.  The model belongs to the K1 class. The $\mathbb{Z}$ topological number is given by $W= \frac{1}{2\pi i}\int _0 ^{2\pi} dk\partial_kln(e^{ik})=1$, which indicates the system belonging to a topological non-trivial class. It is straightforward to check that $\Psi=const$ is an eigenfunction of the Hamiltonian with the eigenvalue $1$. For this case, we do not observe the correspondence between the zero mode and $\mathbb{Z}$ topological number.

            \subsection{Point defects in 2D PC1 class}
           Consider a two-dimensional (2D) Hamiltonian described by
            \begin{eqnarray}
             H(\hat{\bm{k}},\hat{\bm{r}})&=& v \hat{k}_x\gamma_1+ v\hat{k}_y \gamma_2 +  (\Delta _1+i\alpha_1)\gamma_3+  (\Delta_2+i\alpha_2)\gamma_4    \nonumber  \\
                               &=&\left[ \begin{array}{cc}
                                {\bf \bm{\sigma} \cdot k} & \Delta +i\alpha \\
                                  \Delta ^*+i\alpha^*& -{\bf \bm{\sigma} \cdot k}
                                \end{array}
                                \right ],  \label{pc1}
            \end{eqnarray}
           where $(\gamma_1,\gamma_2,\gamma_3,\gamma_4)=(\tau_z\sigma_x,\tau_z\sigma_y,\tau_x,-\tau_y)$, ${\bf k}=(\hat{k}_x,\hat{k}_y)$ and ${\bm \sigma}=(\sigma_x,\sigma_y)$. $\sigma_{x,y,z}$ and $\tau_{x,y,z}$ are Pauli matrices in the spin and particle-hole spaces. Here $\Delta_1$ and $\Delta_2$ are functions of $\bm{\hat{r}}=(\hat{x},\hat{y})$, $\alpha_1$ and $\alpha_2$ are real constants,  $\Delta_1+i\Delta_2=\Delta=|\Delta|e^{i\phi}$ and $\alpha_1+i\alpha_2=\alpha$.  We take $|\alpha_1| \ll |\Delta_1|$, $|\alpha_2| \ll |\Delta_2|$ on $S^1$, where $S^1$ is the loop around the defect. When $\alpha_1=\alpha_2=0$, this model reduces to the Jackiw-Rossi model \cite{JRD2,Teo}.
            The Hamiltonian has $P$ and $C$ symmetries with $\epsilon_c=-1$, $\eta_c=1$, $c=\tau_y\sigma_y$ and $p=\tau_z\sigma_z$, and thus belongs to the $PC1$ class. The topological defect is characterized by a $\mathbb{Z}$ topological number. When $\alpha_1$ and $\alpha_2$ continuous change from 0 to nonzero, the $det(U(\bm{k},\bm{r}))\ne 0$ on $S^1$ in Eq.(\ref{PC1}) because $|\alpha_1| \ll |\Delta_1|$ and $|\alpha_2| \ll |\Delta_2|$ on $S^1$. Then the topological number is same as that of Jackiw-Rossi model with
            \begin{equation}
            n=\frac{1}{2\pi}\int_{s^1}d\phi.
            \end{equation}

            When $\Delta_1=b\hat{x}$ and $\Delta_2=b\hat{y}$ ($b$ is a real constant), 
            in momentum representation, the Hamiltonian takes the following form:
            \begin{equation}
                H(\bm{k},i\partial_{\bm{k}})=v\gamma_1 k_x+ v\gamma_2 k_y+ \gamma_3 (bi\partial_{kx}+i\alpha_1)+ \gamma_4 (bi\partial_{ky}+i\alpha_2).
            \end{equation}
              Define $H'=e^{\frac{\alpha_1 k_x+\alpha_2 k_y}{b}}He^{-\frac{\alpha_1 k_x+\alpha_2 k_y}{b}}$, then $H'$ is given by
              \begin{equation}
                  H'(\hat{\bm{k}},\hat{\bm{r}})=v\gamma_1 \hat{k}_x+ v\gamma_2 \hat{k}_y+ \gamma_3 b \hat{x}+ \gamma_4 b \hat{y},
              \end{equation}
             which is a special form of Jackiw-Rossi Hamiltonian and has a topologically protected zero mode at the defect. Denote the zero mode state of $H'$ as $\Phi '_0(\bm{r})= \langle \bm{r}|\Phi '_0 \rangle$, then $H$ has a topologically protected zero mode $ \Phi_0(\bm{r})=e^{i\alpha_1 \partial_x /b +i \alpha_2 \partial_y/b }\Phi '_0(\bm{r})$ at the defect.

             When we take $\Delta _1=b(\hat{x}^2-\hat{y}^2)$ and $\Delta_2=2b\hat{x}\hat{y}$, the corresponding topological number is given by $n=2$. We can prove that there are two topologically protected zero modes at the defect by a similar method.
            In the momentum space, we can write the Hamiltonian explicitly as
                \begin{equation}
                  \begin{split}
                  H(\bm{k},i\partial_{\bm{k}})=&v\gamma_1 k_x+ v\gamma_2 k_y+ \gamma_3 [b(-\partial _{kx}^2+\partial_{ky}^2)  +i\alpha_1]\\
                                   &+ \gamma_4 (-2b\partial_{kx}\partial_{ky}+i\alpha_2).
                \end{split}
                \end{equation}
                Define $H'=e^{\sqrt{i}(\beta_1 k_x+\beta_2 k_y)}He^{-\sqrt{i}(\beta_1 k_x+\beta_2 k_y)}$, where $\beta_1$ and $\beta_2$ fulfill that: $-b\beta _1^2+b\beta_2^2+\alpha _1=0$ and $ -2b\beta_1\beta_2+\alpha_2=0$, then $H'$ is given by
                \begin{equation}
                  H'(\hat{\bm{k}},\hat{\bm{r}}) =v\gamma_1  \hat{k}_x+ v\gamma_2  \hat{k}_y+ \gamma_3 b( \hat{x}^2- \hat{y}^2) + \gamma_4 2 b  \hat{x} \hat{y} .
                \end{equation}
                The Hermitian Hamiltonian $H'$ has two zero modes, then $H$ also has two zero modes at the defects. The study can be directly extended to situations with $n>2$ . There is a correspondence between the zero modes of defects and the topological number in the 2D $PC1$ class.

                For the non-Hermitian continuous model described by
          \begin{eqnarray}
            H(\bm{k},\bm{r}) &=& \gamma_1(-iv\partial_x+i\beta_1)+ \gamma_2 (-iv\partial_y+i\beta_2) \nonumber \\
            & & + b\Delta_1\gamma_3 + b\Delta_2\gamma_4 , \label{2dopc1}
          \end{eqnarray}
          where $\Delta_1$ and $\Delta_2$ are functions of $\bm{r}$, $\Delta=\Delta_1+i\Delta_2$ has non-trivial winding on $S^1$, and $S^1$ is the loop around the defect. we can make a similarity transformation  $H'=e^{-\frac{\beta_1x+\beta_2y}{v}}H e^{\frac{\beta_1x+\beta_2y}{v}}$ and get
          \begin{equation}
              H'(\hat{\bm{k}},\hat{\bm{r}})=v\hat{k}_x\gamma_1+ v\hat{k}_y\gamma_2 + b\Delta_1\gamma_3 + b\Delta_2\gamma_4 y,
            \end{equation}
   which is a standard Jackiw-Rossi Hamiltonian. Similarly,  we conclude that $H$ has a topologically protected zero mode $\Phi_0(\bm{r})=e^{\frac{\beta_1x+\beta_2y}{v}} \Phi '_0(\bm{r})$, where $\Phi '_0(\bm{r})$ is the zero mode wavefunction of $H'$.

                \subsection{Point defects in 2D C3 class}
                Consider the Hamiltonian:
                \begin{equation}  \label{c3}
                    \begin{split}
                 H(\hat{\bm{k}},\hat{\bm{r}})=& v \hat{k}_x \gamma_1+ v \hat{k}_y \gamma_2+  (\Delta _1+i\alpha_1)\gamma_3+  (\Delta_2+i\alpha_2)\gamma_4 \\
                                    &+h\sigma_z-\mu \tau_z  \\
                                    =&\left[ \begin{array}{cc}
                                        {\bf \bm{\sigma} \cdot k}+\sigma_zh-\mu & \Delta +i\alpha \\
                                          \Delta ^*+i\alpha^*& -{\bf \bm{\sigma} \cdot k}+\sigma_zh+\mu
                                        \end{array}
                                        \right ],
                    \end{split}
                \end{equation}
                where the first four terms are the same as Eq.(\ref{pc1}), $h$ is a  magnetic field and $\mu$ is a chemical potentia. When $\alpha_1=\alpha_2=0$, this model is Fu-Kane model \cite{Herbut,Nishida,FuKane}.
                The Hamiltonian has $C$ symmetries with $\epsilon_c=-1$, $\eta_c=1$ and $c=\tau_y\sigma_y$ and belongs to the $C3$ class. The $h$ and $\mu$ terms couple the zero modes that we discussed in the above section. The topological properties of the system are characterized by a $\mathbb{Z}_2$ topological number. Consider $\alpha_1$ and $\alpha_2$ continuously changing from 0 to nonzero. In this process, $det(H(\bm{k},\bm{r}))\ne 0$ on $S^1$ (loop round the defect) because $|\alpha_1| \ll |\Delta_1|$ and $|\alpha_2| \ll |\Delta_2|$ on $S^1$. Then the topological number is the same as the Hermitian case \cite{Herbut,Nishida} given by
               \begin{equation}
                n=\frac{1}{2\pi}\int_{s^1}d\phi ~ \text{mod} ~ 2.
                \end{equation}

                When $\Delta_1=b\hat{x}$ and $\Delta_2=b\hat{y}$ ($b$ is a real constant), the topological invariant is given by $n=1$. Next, we demonstrate the existence of a zero mode at the defect.
                In momentum representation, the Hamiltonian is written as
                \begin{equation}
                    \begin{split}
                    H(\bm{k},i\partial_{\bm{k}})=&v\gamma_1 k_x+ v\gamma_2 k_y+ \gamma_3 (bi\partial_{kx}+i\alpha_1)+ \\
                                     &\gamma_4 (bi\partial_{ky}+i\alpha_2)+h\sigma_z-\mu \tau_z .
                    \end{split} .
                \end{equation}
                  Define $H'=e^{\frac{\alpha_1 k_x+\alpha_2 k_y}{b}}He^{-\frac{\alpha_1 k_x+\alpha_2 k_y}{b}}$, then $H'$ is given by
                  \begin{equation}
                      H'(\hat{\bm{k}},\hat{\bm{r}})=v\gamma_1 \hat{k}_x+ v\gamma_2 \hat{k}_y+ \gamma_3 b \hat{x}+ \gamma_4 b \hat{y}+h\sigma_z-\mu \tau_z,
                  \end{equation}
                  which is a special form of Fu-Kane Hamiltonian and has a topological protected zero mode at the defect. Denote the zero mode state of the Fu-Kane Hamiltonian $H'$ as $\Phi '_1(\bm{r})=\langle \bm{r}|\Phi '_1 \rangle$, then $H$ has a topological protected zero mode $\Phi_1(\bm{r})=
                  e^{i\alpha_1 \partial_x /b +i \alpha_2 \partial_y/b } \Phi '_1(\bm{r})$ at the defect. When we choose $\Delta _1=b(\hat{x}^2-\hat{y}^2)$ and $\Delta_2=2b\hat{x}\hat{y}$, the corresponding topological number is $n=0$. And we can demonstrate that there is no topologically protected zero mode at the defect by a similar method. There is a correspondence between the zero mode of defect and the topological number in the 2D $C3$ class.

                  For the non-Hermitian continuous model described by
                  \begin{equation}       \label{2dopc3}
                    \begin{split}
                    H(\bm{k},\bm{r})=&\gamma_1(-iv\partial_x+i\beta_1)+ \gamma_2 (-iv\partial_y+i\beta_2)+ b\Delta_1\gamma_3 \\
                      &+b\Delta_2\gamma_4+h\sigma_z-\mu \tau_z ,
                  \end{split}
                  \end{equation}
                  where $\Delta_1$ and $\Delta_2$ are functions of $\bm{r}$, $\Delta=\Delta_1+i\Delta_2$ has odd winding number on $S^1$, and $S^1$ is the loop around the defect. We can prove that there is topologically protected zero mode by a similar method.

                  \subsection{Point defects in 2D PQC1 and QC5 class}
             Consider the Hamiltonian Eq.(\ref{pc1}) with $\alpha_1=\alpha_2=0$, i.e., the Jackiw-Rossi model. This model has one more symmetry $Q$ than Eq.(\ref{pc1}) with $q=\mathbb{I}$. According to the BL classification, the Jackiw-Rossi model belongs to the PQC1 class, and the Hermitian model has a correspondence between zero modes and $\mathbb{Z}$ topological number. Similarly, consider the Hamiltonian Eq.(\ref{c3}) with $\alpha_1=\alpha_2=0$, i.e., the Fu-Kane model.  This model has one more symmetry $Q$ than Eq.(\ref{c3}) with $q=\mathbb{I}$.  According to the BL classification, the Fu-Kane model belongs to the  QC5 class, and the Hermitian model has a correspondence between zero modes and $\mathbb{Z}_2$ topological number.

                   \section{Summary}
                     In summary, we have studied topological defects in non-Hermitian systems with point gap, real gap and imaginary gap and made a topological classification
                   for all the BL or GBL classes in all dimensions. While the BL class covers all 38  nonequivalent symmetry classes for the point-gap systems, we find that a full classification for non-Hermitian systems with line gap should include 54 nonequivalent GBL classes, which are a natural generalization of BL classes.
                   The periodical classification tables of point gap defects are summarized in table II for the AZ classes and in table III for the BL classes with point gap, respectively, and periodical classification tables of real and imaginary gap defects are summarized in table IV and V (GBL class), respectively.
                  By considering some concrete examples of point gap defects, we constructed explicitly the topological invariants.  Through the study of some concrete models, we calculated explicitly the topological invariants and discussed the correspondence between topological invariants and zero modes at the defect for some classes.

\begin{acknowledgments}
The work is supported by NSFC under Grants No. 11425419 and the National Key Research and Development Program of China (2016YFA0300600 and 2016YFA0302104).
\end{acknowledgments}

                     \section{appendix}
                     \subsection{Real gap classification}
                     According to the discussion in the main text, the classification  is equivalent to the classification of a Hermitian Hamiltonian with corresponding symmetries:
                     \begin{align}
                      H=\epsilon_k kH^*k^{-1}&, kk^*=\eta_k \mathbb{I}&K \textrm{ sym.} \label{eq:syms1} \\
                      H=\epsilon_q qH q^{-1}&, q^2=\mathbb{I}&Q \textrm{ sym.}\label{eq:syms2}\\
                      H=\epsilon_c cH^* c^{-1}&, cc^*=\eta_c \mathbb{I}&C \textrm{ sym.}\label{eq:syms3}\\
                      H=-pHp^{-1}&, p^2=\mathbb{I}&P \textrm{ sym.}\label{eq:syms4}
                      \end{align}
                         We can use Clifford algebra to represent the Hamiltonian:
                         \begin{equation}
                             H({\bf k,r})=\gamma_0+k_1\gamma_1^k+...+k_d\gamma_d^k+r_1\gamma_1^r+...+r_D\gamma_D^r .
                         \end{equation}
                         Define $K=k\mathcal{K}$, $Q=q$, $C=c\mathcal{K}$ and $P=p$, where $\mathcal{K}$ is a complex conjugate operator. We can get the space of mass term by constructing the Clifford algebra's extension. The correspondence between Clifford algebra's extension and space of mass term was listed in Table I and Table VI.
                          \begin{table}[htbp]
                            \caption{\label{tab:table7} The correspondence between Clifford algebra's extension $Cl_{p,q+p}\otimes Cl_{0,m}\rightarrow Cl_{p,p+q+1}\otimes Cl_{0,m}$($Cl_{q+p-2,p}\otimes Cl_{0,m}\rightarrow Cl_{q+p-1,p}\otimes Cl_{0,m}$) and the space of mass term.}
                          \begin{ruledtabular}
                          \begin{tabular}{ccc}
                            m(mod 8)&Space of mass term&$\pi_0$\\
                            \hline
                            0&$R_{q}$&$\pi_0$($R_{q}$)\\
                            1&$R_{q}\times R_{q}$&$\pi_0$($R_{q}$)$\oplus\pi_0$($R_{q}$)\\
                            2&$R_{q}$&$\pi_0$($R_{q}$)\\
                            3&$C_{q}$&$\pi_0$($C_{q}$)\\
                            4&$R_{q+4}$&$\pi_0$($R_{q+4}$)\\
                            5&$R_{q+4}\times R_{q+4}$&$\pi_0$($R_{q+4}$)$\oplus\pi_0$($R_{q+4}$)\\
                            6&$R_{q+4}$&$\pi_0$($R_{q+4}$)\\
                            7&$R_{q+4}$&$\pi_0$($R_{q+4}$)\\
                            \end{tabular}
                         \end{ruledtabular}
                        \end{table}
                          Then we get the topological classification by calculating the homotopy group of the space of mass term. The following is the topological classification for some GBL classes.

                          {\bf Class Non:} The generators are $\{\gamma_0,\gamma_1^k,\gamma_2^k,...,\gamma_d^k,\gamma_1^r,\gamma_2^r,...,\gamma_D^r\}$. Clifford algebra's extension is $\{\gamma_1^k,\gamma_2^k,...,\gamma_d^k,\gamma_1^r,\gamma_2^r,...,\gamma_D^r \}\rightarrow  \{\gamma_0,\gamma_1^k,\gamma_2^k,...,\gamma_d^k,\gamma_1^r,\gamma_2^r,...,\gamma_D^r\}=Cl_{d+D}\rightarrow Cl_{d+D+1}$. The space of mass term is $C_{d+D}=C_{-(d-D)}=C_{-\delta}$. The topological classification is determined by $\pi_0(C_{-\delta})=\mathbb{Z}~(0)$ for even (odd) $\delta$.

                          {\bf Class P:} The generators are $\{\gamma_0,\gamma_1^k,\gamma_2^k,...,\gamma_d^k,\gamma_1^r,\gamma_2^r,...,\gamma_D^r,P\}$. Clifford algebra's extension is $\{\gamma_1^k,\gamma_2^k,...,\gamma_d^k,\gamma_1^r,\gamma_2^r,...,\gamma_D^r,P \}\rightarrow  \{\gamma_0,\gamma_1^k,\gamma_2^k,...,\gamma_d^k,\gamma_1^r,\gamma_2^r,...,\gamma_D^r,P\}=Cl_{d+D+1}\rightarrow Cl_{d+D+2}$. The space of mass term is $C_{d+D+1}=C_{-(d-D)+1}=C_{-\delta+1}$. The topological classification is determined by $\pi_0(C_{-\delta+1})=0~(\mathbb{Z})$ for even (odd) $\delta$.

                          {\bf Class Qa:} The generators are $\{\gamma_0,\gamma_1^k,\gamma_2^k,...,\gamma_d^k,\gamma_1^r,\gamma_2^r,...,\gamma_D^r,Q\}$. Clifford algebra's extension is $\{\gamma_1^k,\gamma_2^k,...,\gamma_d^k,\gamma_1^r,\gamma_2^r,...,\gamma_D^r\}\otimes\{ Q \} \rightarrow  \{\gamma_0,\gamma_1^k,\gamma_2^k,...,\gamma_d^k,\gamma_1^r,\gamma_2^r,...,\gamma_D^r\}\otimes\{ Q \}=Cl_{d+D}\otimes Cl_1 \rightarrow Cl_{d+D+1}\otimes Cl_1 $. The space of mass term is $C_{d+D}\times C_{d+D}=C_{-\delta}\times C_{-\delta}$. The topological classification is determined by $\pi_0(C_{-\delta}\times C_{-\delta})=\mathbb{Z}\oplus \mathbb{Z}~(0\oplus 0)$ for even (odd) $\delta$.

                          {\bf Class C1:} The generators are $\{\gamma_0,\gamma_1^k,\gamma_2^k,...,\gamma_d^k,\gamma_1^r,\gamma_2^r,...,\gamma_D^r,C,J\}$. Clifford algebra's extension is $\{\gamma_1^k,\gamma_2^k,...,\gamma_d^k,J\gamma_1^r,J\gamma_2^r,...,J\gamma_D^r,C,JC\} \rightarrow  \{J\gamma_0,\gamma_1^k,\gamma_2^k,...,\gamma_d^k,J\gamma_1^r,J\gamma_2^r,...,J\gamma_D^r,C,JC\}=Cl_{D,d+2} \rightarrow Cl_{D+1,d+2} $. The space of mass term is $R_{D-d}=R_{-\delta}$.

                          {\bf Class PQ1:} The generators are $\{\gamma_0,\gamma_1^k,\gamma_2^k,...,\gamma_d^k,\gamma_1^r,\gamma_2^r,...,\gamma_D^r,P,Q\}$. Clifford algebra's extension is $\{\gamma_1^k,\gamma_2^k,...,\gamma_d^k,\gamma_1^r,\gamma_2^r,...,\gamma_D^r,P\}\otimes\{ Q \} \rightarrow  \{\gamma_0,\gamma_1^k,\gamma_2^k,...,\gamma_d^k,\gamma_1^r,\gamma_2^r,...,\gamma_D^r,P\}\otimes\{ Q \}=Cl_{d+D+1}\otimes Cl_1 \rightarrow Cl_{d+D+2}\otimes Cl_1 $. The space of mass term is $C_{d+D+1}\times C_{d+D+1}=C_{-\delta+1}\times C_{-\delta+1}$. The topological classification is determined by $\pi_0(C_{-\delta+1}\times C_{-\delta+1})=\mathbb{Z}\oplus \mathbb{Z} ~( 0\oplus 0 )$ for odd (even) $\delta$.

                          {\bf Class PK1:} The generators are $\{\gamma_0,\gamma_1^k,\gamma_2^k,...,\gamma_d^k,\gamma_1^r,\gamma_2^r,...,\gamma_D^r,P,K,J\}$. Clifford algebra's extension is $\{\gamma_1^k,\gamma_2^k,...,\gamma_d^k,J\gamma_1^r,J\gamma_2^r,...,J\gamma_D^r,JP,K,JK\} \rightarrow  \{J\gamma_0,\gamma_1^k,\gamma_2^k,...,\gamma_d^k,J\gamma_1^r,J\gamma_2^r,...,J\gamma_D^r,JP,K,JK\}=Cl_{D+1,d+2} \rightarrow Cl_{D+2,d+2} $. The space of mass term is $R_{D-d+1}=R_{1-\delta}$.

                          {\bf Class PC1:} The generators are $\{\gamma_0,\gamma_1^k,\gamma_2^k,...,\gamma_d^k,\gamma_1^r,\gamma_2^r,...,\gamma_D^r,P,C,J\}$. Clifford algebra's extension is $\{\gamma_1^k,\gamma_2^k,...,\gamma_d^k,J\gamma_1^r,J\gamma_2^r,...,J\gamma_D^r,JP,C,JC\} \rightarrow  \{J\gamma_0,\gamma_1^k,\gamma_2^k,...,\gamma_d^k,J\gamma_1^r,J\gamma_2^r,...,J\gamma_D^r,JP,C,JC\}=Cl_{D+1,d+2} \rightarrow Cl_{D+2,d+2} $. The space of mass term is $R_{D-d+1}=R_{1-\delta}$.

                          {\bf Class QC1a:} The generators are $\{\gamma_0,\gamma_1^k,\gamma_2^k,...,\gamma_d^k,\gamma_1^r,\gamma_2^r,...,\gamma_D^r,Q,C,J\}$. Clifford algebra's extension is $\{\gamma_1^k,\gamma_2^k,...,\gamma_d^k,J\gamma_1^r,J\gamma_2^r,...,J\gamma_D^r,C,JC\}\otimes \{Q \} \rightarrow  \{J\gamma_0,\gamma_1^k,\gamma_2^k,...,\gamma_d^k,J\gamma_1^r,J\gamma_2^r,...,J\gamma_D^r,C,JC\}\otimes \{Q \}=Cl_{D,d+2}\times Cl_{0,1} \rightarrow Cl_{D+1,d+2}\times Cl_{0,1} $. The space of mass term is $R_{D-d}\times R_{D-d}=R_{-\delta}\times R_{-\delta}$.

                          {\bf Class PQC1:} The generators are $\{\gamma_0,\gamma_1^k,\gamma_2^k,...,\gamma_d^k,\gamma_1^r,\gamma_2^r,...,\gamma_D^r,P,Q,C,J\}$. Clifford algebra's extension is $\{\gamma_1^k,\gamma_2^k,...,\gamma_d^k,J\gamma_1^r,J\gamma_2^r,...,J\gamma_D^r,JP,C,JC\}\otimes \{Q \} \rightarrow  \{J\gamma_0,\gamma_1^k,\gamma_2^k,...,\gamma_d^k,J\gamma_1^r,J\gamma_2^r,...,J\gamma_D^r,JP,C,JC\}\otimes \{Q \}=Cl_{D+1,d+2}\times Cl_{0,1} \rightarrow Cl_{D+2,d+2}\times Cl_{0,1} $. The space of mass term is $R_{D-d+1}\times R_{D-d+1}=R_{1-\delta}\times R_{1-\delta}$.

                     \subsection{Imaginary gap classification}
                      According to the discussion in the main text, the classification  is equivalent to the classification of a Hermitian Hamiltonian with corresponding symmetries:
                     \begin{align}
                        H({\bf -k,r})=-\epsilon_k kH({\bf k,r})^*k^{-1}&, kk^*=\eta_k \mathbb{I}&K \textrm{ sym.} \label{eq:syms1} \\
                        H({\bf k,r})=-\epsilon_q qH({\bf k,r}) q^{-1}&, q^2=\mathbb{I}&Q \textrm{ sym.}\label{eq:syms2}\\
                        H({\bf -k,r})=\epsilon_c cH({\bf k,r})^*c^{-1}&, cc^*=\eta_c \mathbb{I}&C \textrm{ sym.}\label{eq:syms3}\\
                        H({\bf k,r})=-pH({\bf k,r})p^{-1}&, p^2=\mathbb{I}&P \textrm{ sym.}\label{eq:syms4}
                        \end{align}
                        We can use Clifford algebra to represent the Hamiltonian:
                        \begin{equation}
                            H(k,r)=\gamma_0+k_1\gamma_1^k+...+k_d\gamma_d^k+r_1\gamma_1^r+...+r_D\gamma_D^r .
                        \end{equation}
                        Define $K=k\mathcal{K}$, $Q=q$, $C=c\mathcal{K}$ and $P=p$, where $\mathcal{K}$ is the complex conjugate operator. Then we can get the space of mass term by constructing the Clifford algebra's extension. And we get the topological classification by calculating the homotopy group of the space of mass term. The $C$ and $P$ give the same symmetry constraint for the topological equivalent Hermitian Hamiltonian in real-gap systems and imaginary-gap systems. Then the two systems have the same topological classification for GBL classes: $Non$, $P$, $C1-4$ and $PC1-4$. For one of  PQC1-8 classes, the constraint of topologically equivalent Hermitian Hamiltonian in imaginary-gap systems transforms to that in the corresponding real-gap systems after defining $\tilde{q}=\sqrt{\epsilon_{pq}}pq$. Then real-gap and imaginary-gap classification of class PQC1-8 are also same. The following is the topological classification for some GBL classes.

                         {\bf Class Qa:}  The generators are $\{\gamma_0,\gamma_1^k,\gamma_2^k,...,\gamma_d^k,\gamma_1^r,\gamma_2^r,...,\gamma_D^r,Q\}$. Clifford algebra's extension is $\{\gamma_1^k,\gamma_2^k,...,\gamma_d^k,\gamma_1^r,\gamma_2^r,...,\gamma_D^r,Q \}\rightarrow  \{\gamma_0,\gamma_1^k,\gamma_2^k,...,\gamma_d^k,\gamma_1^r,\gamma_2^r,...,\gamma_D^r,Q\}=Cl_{d+D+1}\rightarrow Cl_{d+D+2}$. The space of mass term is $C_{d+D+1}=C_{-(d-D)+1}=C_{-\delta+1}$. The topological classification is determined by $\pi_0(C_{-\delta+1})=0 ~(\mathbb{Z})$ for even (odd) $\delta$.

                         {\bf Class K1a:}  The generators are $\{\gamma_0,\gamma_1^k,\gamma_2^k,...,\gamma_d^k,\gamma_1^r,\gamma_2^r,...,\gamma_D^r,K,J\}$. Clifford algebra's extension is $\{J\gamma_1^k,J\gamma_2^k,...,J\gamma_d^k,\gamma_1^r,\gamma_2^r,...,\gamma_D^r,K,JK\} \rightarrow  \{\gamma_0,J\gamma_1^k,J\gamma_2^k,...,J\gamma_d^k,\gamma_1^r,\gamma_2^r,...,\gamma_D^r,K,JK\}=Cl_{d,D+2} \rightarrow Cl_{d,D+3} $. The space of mass term is $R_{D-d+2}=R_{2-\delta}$.

                         {\bf Class PQ1:}  The generators are $\{\gamma_0,\gamma_1^k,\gamma_2^k,...,\gamma_d^k,\gamma_1^r,\gamma_2^r,...,\gamma_D^r,P,Q\}$. Clifford algebra's extension is $\{\gamma_1^k,\gamma_2^k,...,\gamma_d^k,\gamma_1^r,\gamma_2^r,...,\gamma_D^r,P\}\otimes \{PQ \}\rightarrow  \{\gamma_0,\gamma_1^k,\gamma_2^k,...,\gamma_d^k,\gamma_1^r,\gamma_2^r,...,\gamma_D^r,P\}\otimes \{PQ\}=Cl_{d+D+1}\times Cl_{1}\rightarrow Cl_{d+D+2}\times Cl_{1}$. The space of mass term is $C_{d+D+1}\times C_{d+D+1}=C_{1-\delta}\times C_{1-\delta}$.

                         {\bf Class PK1:}  The generators are $\{\gamma_0,\gamma_1^k,\gamma_2^k,...,\gamma_d^k,\gamma_1^r,\gamma_2^r,...,\gamma_D^r,P,K,J\}$. Clifford algebra's extension is $\{J\gamma_1^k,J\gamma_2^k,...,J\gamma_d^k,\gamma_1^r,\gamma_2^r,...,\gamma_D^r,JP,K,JK\} \rightarrow  \{\gamma_0,J\gamma_1^k,J\gamma_2^k,...,J\gamma_d^k,\gamma_1^r,\gamma_2^r,...,\gamma_D^r,JP,K,JK\}=Cl_{d+1,D+2} \rightarrow Cl_{d+1,D+3} $. The space of mass term is $R_{D-d+1}=R_{1-\delta}$.

                         {\bf Class PC1:}  The generators are $\{\gamma_0,\gamma_1^k,\gamma_2^k,...,\gamma_d^k,\gamma_1^r,\gamma_2^r,...,\gamma_D^r,P,C,J\}$. Clifford algebra's extension is $\{\gamma_1^k,\gamma_2^k,...,\gamma_d^k,J\gamma_1^r,J\gamma_2^r,...,J\gamma_D^r,JP,C,JC\} \rightarrow  \{J\gamma_0,\gamma_1^k,\gamma_2^k,...,\gamma_d^k,J\gamma_1^r,J\gamma_2^r,...,J\gamma_D^r,JP,C,JC\}=Cl_{D+1,d+2} \rightarrow Cl_{D+2,d+2} $. The space of mass term is $R_{D-d+1}=R_{1-\delta}$.

                         {\bf Class QC1a:}  The generators are $\{\gamma_0,\gamma_1^k,\gamma_2^k,...,\gamma_d^k,\gamma_1^r,\gamma_2^r,...,\gamma_D^r,Q,C,J\}$. Clifford algebra's extension is $\{\gamma_1^k,\gamma_2^k,...,\gamma_d^k,J\gamma_1^r,J\gamma_2^r,...,J\gamma_D^r,JQ,C,JC\} \rightarrow  \{J\gamma_0,\gamma_1^k,\gamma_2^k,...,\gamma_d^k,J\gamma_1^r,J\gamma_2^r,...,J\gamma_D^r,JQ,C,JC\}=Cl_{D+1,d+2} \rightarrow Cl_{D+2,d+2} $. The space of mass term is $R_{D-d+1}=R_{1-\delta}$.

                          \subsection{The correspondence between different notations of BL class and GBL class}
                          There are three different notations of Bernard-LeClair class: Kawabata-Shiozaki-Ueda-Sato (KSUS) notations \cite{Sato}, Zhou-Lee (ZL) notations \cite{Zhou} and our (LC) notations used in the present work. For the convenience of comparing with the existing results in references, we list the correspondence between different notations of Bernard-LeClair class in table VII. For the line gap systems, there are two different notations of generalized Bernard-LeClair class: Kawabata-Shiozaki-Ueda-Sato (KSUS) notations and our (LC) notations. We list the correspondence of different notations of generalized Bernard-LeClair class in table VIII.

 \onecolumngrid \clearpage

    \begin{table*}[htbp]\footnotesize
      \begin{center}
      \caption{\label{tab:table4}  The correspondence between different notations of Bernard-LeClair class. }
    \begin{tabular}{|c|c|c|c|c|c|}
        \hline
        LC & ZL &  KSUS & LC & ZL &  KSUS \\
      \hline
    Non&1&A & QC2& 18& CI$^\dagger$, $\eta_-$AII\\
      \hline
    P&2&$\mathcal{S}$A&QC3& 15 & CII$^\dagger$, $\eta_+$AII\\
      \hline
    Q&  3 &AIII, $\eta$A&QC4& 19& DIII$^\dagger$, $\eta_-$AI\\
      \hline
    K1& 34& AI, D$^\dagger$&QC5& 16& BDI, $\eta_+$D\\
      \hline
    K2& 35 & AII, C$^\dagger$&QC6& 20& DIII, $\eta_-$D \\
      \hline
    C1& 6 & AI$^\dagger$&QC7& 17& CII, $\eta_+$C \\
      \hline
    C2& 7 & AII$^\dagger$&QC8& 21& CI, $\eta_-$C\\
      \hline
    C3& 8 & D &PQC1& 22&$\mathcal{S_{++}}$BDI, $\eta_{++}$BDI\\
      \hline
    C4& 9& C& PQC2& 32&$\mathcal{S_{++}}$DIII, $\eta_{--}$DIII\\
      \hline
    PQ1& 4 & $\mathcal{S_+}$AIII, $\eta_+$AIII &PQC3& 28 &$\mathcal{S_{+-}}$CI, $\eta_{+-}$CI\\
    \hline
    PQ2& 5& $\mathcal{S_-}$AIII, $\eta_-$AIII&PQC4& 31&$\mathcal{S_{+-}}$CII, $\eta_{-+}$CII\\
    \hline
    PK1& 36& $\mathcal{S_+}$AI&PQC5& 23&$\mathcal{S_{++}}$CII, $\eta_{++}$CII\\
    \hline
    PK2& 37& $\mathcal{S_+}$AII&PQC6& 33&$\mathcal{S_{++}}$CI, $\eta_{--}$CI\\
    \hline
   PK3& 38& $\mathcal{S_-}$AI, $\mathcal{S_-}$AII &PQC7&29&$\mathcal{S_{+-}}$DIII, $\eta_{+-}$DIII\\
      \hline
    PC1& 10&$\mathcal{S_+}$D&PQC8& 30&$\mathcal{S_{+-}}$BDI, $\eta_{-+}$BDI\\
      \hline
    PC2& 12&$\mathcal{S_-}$C&PQC9& 26&$\mathcal{S_{--}}$CI, $\mathcal{S_{--}}$CII, $\eta_{++}$CI, $\eta_{--}$CII\\
      \hline
    PC3& 11&$\mathcal{S_+}$C&PQC10& 24&$\mathcal{S_{-+}}$BDI, $\mathcal{S_{-+}}$DIII, $\eta_{+-}$BDI, $\eta_{-+}$DIII\\
      \hline
    PC4& 13&$\mathcal{S_-}$D&PQC11& 27&$\mathcal{S_{--}}$DIII, $\mathcal{S_{--}}$BDI, $\eta_{++}$DIII, $\eta_{--}$BDI\\
      \hline
    QC1& 14&BDI$^\dagger$, $\eta_+$AI&PQC12& 25&$\mathcal{S_{-+}}$CII, $\mathcal{S_{-+}}$CI, $\eta_{+-}$CII, $\eta_{-+}$CI\\
      \hline
  \end{tabular}
  \end{center}
\end{table*}

    \begin{table*}[htbp]\footnotesize
      \begin{center}
      \caption{\label{tab:table4}  The correspondence between different notations of generalized Bernard-LeClair class.  }
    \begin{tabular}{|c|c|c|c|}
        \hline
        LC &  KSUS &LC &  KSUS\\
      \hline
    Non&A   &  QC3b&  CII$^\dagger$\\
      \hline
    P&$\mathcal{S}$A& QC4a&  $\eta_-$AI\\
      \hline
    Qa&   $\eta$A& QC4b&  DIII$^\dagger$\\
      \hline
      Qb&  AIII&QC5a& $\eta_+$D\\
      \hline
    K1a&  AI& QC5b& BDI\\
      \hline
      K1b&  D$^\dagger$&QC6a& $\eta_-$D\\
      \hline
    K2a&  AII&QC6b&  DIII\\
      \hline
      K2b&  C$^\dagger$  &QC7a&   $\eta_+$C\\
      \hline
    C1& AI$^\dagger$ &QC7b&  CII\\
      \hline
    C2& AII$^\dagger$&QC8a& $\eta_-$C\\
      \hline
    C3& D & QC8b&  CI\\
      \hline
    C4&  C&PQC1&$\mathcal{S_{++}}$BDI, $\eta_{++}$BDI\\
      \hline
    PQ1&  $\mathcal{S_+}$AIII, $\eta_+$AIII& PQC2&$\mathcal{S_{++}}$DIII, $\eta_{--}$DIII\\
    \hline
    PQ2&  $\mathcal{S_-}$AIII, $\eta_-$AIII&PQC3& $\mathcal{S_{+-}}$CI, $\eta_{+-}$CI \\
    \hline
    PK1&  $\mathcal{S_+}$AI&PQC4& $\mathcal{S_{+-}}$CII, $\eta_{-+}$CII\\
    \hline
    PK2&  $\mathcal{S_+}$AII&PQC5& $\mathcal{S_{++}}$CII, $\eta_{++}$CII\\
    \hline
   PK3a&  $\mathcal{S_-}$AI&PQC6& $\mathcal{S_{++}}$CI, $\eta_{--}$CI\\
      \hline
   PK3b&  $\mathcal{S_-}$AII&PQC7&$\mathcal{S_{+-}}$DIII, $\eta_{+-}$DIII\\
      \hline
    PC1& $\mathcal{S_+}$D&PQC8& $\mathcal{S_{+-}}$BDI, $\eta_{-+}$BDI\\
      \hline
    PC2& $\mathcal{S_-}$C&PQC9a& $\mathcal{S_{--}}$CI, $\eta_{++}$CI\\
      \hline
    PC3& $\mathcal{S_+}$C&PQC9b&  $\mathcal{S_{--}}$CII, $\eta_{--}$CII\\
      \hline
    PC4& $\mathcal{S_-}$D&PQC10a& $\mathcal{S_{-+}}$BDI $\eta_{+-}$BDI\\
      \hline
    QC1a&  $\eta_+$AI&PQC10b& $\mathcal{S_{-+}}$DIII, $\eta_{-+}$DIII\\
      \hline
      QC1b& BDI$^\dagger$ &PQC11a& $\mathcal{S_{--}}$DIII $\eta_{++}$DIII\\
      \hline

    QC2a&   $\eta_-$AII&PQC11b& $\mathcal{S_{--}}$BDI $\eta_{--}$BDI\\
      \hline
      QC2b&  CI$^\dagger$ &PQC12a& $\mathcal{S_{-+}}$CII, $\eta_{+-}$CII\\
      \hline

    QC3a&   $\eta_+$AII &PQC12b&$\mathcal{S_{-+}}$CI, $\eta_{-+}$CI\\
      \hline

  \end{tabular}
  \end{center}
\end{table*}

\twocolumngrid

\end{document}